# N$_2$-H$_2$ capacitively coupled radio-frequency discharges at low pressure. Part I. Experimental results: effect of the H$_2$ amount on electrons, positive ions and ammonia formation.

**Audrey Chatain*[1,2], Miguel Jiménez-Redondo[3,4], Ludovic Vettier[1], Olivier Guaitella[2], Nathalie Carrasco[1], Luis Lemos Alves[5], Luis Marques[3], Guy Cernogora[1]**

[1]LATMOS/IPSL, UVSQ Université Paris-Saclay, Sorbonne Université, CNRS, 78280 Guyancourt, France
[2]LPP, École Polytechnique, Sorbonne Université, université Paris-sud, CNRS, 91128, Palaiseau, France
[3]Centro de Física das Universidades do Minho e do Porto, Universidade do Minho, 4710-057, Braga, Portugal
[4]Present address: Instituto de Estructura de la Materia, IEM-CSIC, Serrano 123, 28006 Madrid, Spain
[5]Instituto de Plasmas e Fusão Nuclear, Instituto Superior Técnico, Univ. Técnica de Lisboa, Lisboa, Portugal

*e-mail: audrey.chatain@latmos.ipsl.fr

## Abstract

The mixing of N$_2$ with H$_2$ leads to very different plasmas from pure N$_2$ and H$_2$ plasma discharges. Numerous issues are therefore raised involving the processes leading to ammonia (NH$_3$) formation.

The aim of this work is to better characterize capacitively-coupled radiofrequency plasma discharges in N$_2$ with few percents of H$_2$ (up to 5%), at low pressure (0.3 to 1 mbar) and low coupled power (3 to 13 W). Both experimental measurements and numerical simulations are performed. For clarity, we separated the results in two complementary parts. The actual one (first part), presents the details on the experimental measurements, while the second focuses on the simulation, a hybrid model combining a 2D fluid module and a 0D kinetic module.

Electron density is measured by a resonant cavity method. It varies from 0.4 to $5.10^9$ cm$^{-3}$, corresponding to ionization degrees from $2.10^{-8}$ to $4.10^{-7}$. Ammonia density is quantified by combining IR absorption and mass spectrometry. It increases linearly with the amount of H$_2$ (up to $3.10^{13}$ cm$^{-3}$ at 5% H$_2$). On the contrary, it is constant with pressure, which suggests the dominance of surface processes on the formation of ammonia. Positive ions are measured by mass spectrometry. Nitrogen-bearing ions are hydrogenated by the injection of H$_2$, N$_2$H$^+$ being the major ion as soon as the amount of H$_2$ is > 1%. The increase of pressure leads to an increase of secondary ions formed by ion/radical – neutral collisions (ex: N$_2$H$^+$, NH$_4^+$, H$_3^+$), while an increase of the coupled power favours ions formed by direct ionization (ex: N$_2^+$, NH$_3^+$, H$_2^+$).

# 1. Introduction

Nitrogen and hydrogen are usual gases, and a lot of studies have been conducted with the objective to understand pure $N_2$ [1,2] and pure $H_2$ [3,4] plasmas. However, as soon as the two gases are mixed, plasmas become different, and much harder to understand. Today questionings deal with the chemical mechanisms happening in the gas phase and on the surfaces in contact, and especially with the processes leading to the formation of ammonia ($NH_3$).

Many technological applications use $N_2$-$H_2$ plasmas nowadays, as the plasma species formed with both hydrogen and nitrogen have interesting properties. A major issue is the industry of thin film growth, among which are the silicon (SiN) films fundamental for the semiconductor and the photovoltaic industries. These films are formed by plasma-enhanced chemical vapour deposition (PECVD), with gas mixtures containing Si, $N_2$ and $H_2$ [5–8]. Another major industrial application of $N_2$-$H_2$ plasmas is nitriding, used to harden metal surfaces [9,10]. In nuclear fusion, $N_2$ is added to hydrogen plasma to inhibit organic film deposition on walls [11–13]. The formation of ammonia has a strong industrial interest as ammonia is already produced in huge quantities to be used as a basic precursor for the synthesis of chemicals such as fertilizers. The current technique known as the Haber-Bosch process has a yield of 15-20% and research is on-going to exceed this yield using plasmas [14]. In aerospace applications, $N_2$-$H_2$ plasmas are used for propulsion through arcjet thrusters [15]. In all these applications, the understanding of $N_2$-$H_2$ plasma is fundamental to optimize the protocols and designs of technologies.

The understanding of $N_2$-$H_2$ plasmas is not only considered for in the industrial world, but also in astrophysics and planetary sciences. In the interstellar medium for instance, nitrogen and hydrogen excited species and ions are present in large quantity. A current issue is to understand the formation processes of ammonia which abundance is not explained by current models. [16–18] explain it could be due to an additional surface chemistry on ice and dust particles. Another issue is about the detection of $N_2$ in radio astronomy. $N_2$ has no rotational transitions and is therefore not detectable. However, $N_2H^+$ can be detected. To deduce the density of $N_2$ from measurements on $N_2H^+$ requires a complete understanding of the sources and sinks of $N_2H^+$ in this environment [19,20]. The study of the ionospheres of planets also often deals with $N_2$-$H_2$ plasmas. Among them, the ionosphere of Titan, the largest moon of Saturn, is of major interest as it is the place of formation of very complex organic aerosols. Titan's ionosphere is mainly composed of ~97.6% of $N_2$, ~2% of $CH_4$ and ~0.4% of $H_2$ [21]. To understand the complex ion chemistry happening there, different groups simulate the ionosphere of Titan with laboratory cold plasmas, with for instance a radiofrequency capacitively coupled plasma (RF CCP) discharge [22], a pulsed discharge nozzle (PDN) freejet planar expansion [23] or a DC glow discharge [24]. In such plasmas, nitrogen and hydrogen are massively present, and it is necessary to understand their behaviour in a simplified $N_2$-$H_2$ discharge before adding $CH_4$ to the mixture. In particular, $NH_3$ and $NH_4^+$ are present in Titan's upper atmosphere [21,25] but current models deriving $NH_3$ amounts from $NH_4^+$ measurements underestimate $NH_3$ production by a factor of 10-100 [26]. The possible production of $NH_3$ on surfaces (like on solid aerosols) is still to be investigated.

The study of plasmas requires experiments and numerical simulations to closely work together. Diagnosis and modelling of $N_2$-$H_2$ discharges started with the works of [27,28] on DC glow discharges. [29,30] continued this work with the complexification of the chemical model, especially with the introduction of surface kinetics. Ammonia density among other species could not be explained by bulk processes, and led [30] to discover the fundamental role of surface reactions on the global discharge kinetics. They also showed that surface kinetics is highly dependent on the surface state: the surface material, but also the way it has been processed before the measurements. The complex ion processes are studied in [16,17,31] working at low pressure (0.8-8 Pa), while details of mechanism at the surface are described in [14,32] working at atmospheric pressure.

Plasma characteristics depend on the geometry and on the nature of the discharge. All the studies cited above have been performed in DC glow discharges, which are very homogeneous and can be modelled in 0D.

However, $N_2$-$H_2$ plasma applications cited above do not all work in DC glow discharges. In particular, material processing or functionalization if often performed with radiofrequency capacitively coupled (RF CCP) discharges [10,33], as well as the study of Titan's ionosphere in [22]. Some first studies addressed microwave discharges [34] and radiofrequency inductively coupled plasmas [35]. However, to our knowledge, none of them deals with RF CCP discharges.

RF CCP discharges are more complex to diagnose and to model than DC plasmas. Indeed, charged species are not homogeneously distributed in the plasma. For that purpose, [36,37] implemented a two-dimensional model to describe pure $H_2$ RF CCP discharges. It is a hybrid model in which the dynamics of charged species are described by a 2D, time-dependent fluid module, while a zero-dimensional kinetic module solves the chemistry. This work has been continued for pure nitrogen in [38].

Experiments and models in glow DC discharges (references cited above) show the strong modifications of pure $N_2$ (and pure $H_2$) plasmas with the small addition of $H_2$ (resp. $N_2$), and the major effect of surface processes on the plasma global kinetics.

Our objective here is to analyse such modifications in capacitively coupled radio-frequency discharges by the addition of small amounts of $H_2$ (up to 5%) in an initially pure $N_2$ plasma, at low pressure (0.3 – 1 mbar) and low power (3 – 13 W) conditions.

Different measurements are performed, probing electric parameters, electron density, ammonia density and positive ion evolutions. These measurements have then been used to improve the model previously adapted for pure $N_2$ discharges [38], the main additions being the new $N_xH_y$ species, their gas-phase reactions and the surface processes. For clarity, we divided the work in two parts: paper I (this one) focusing on experimental measurements and paper II [39] describing the numerical simulation and the comparison with the experimental data.

## 2. Experimental setup

### *2.1 A RF CCP discharge in $N_2$-$H_2$*

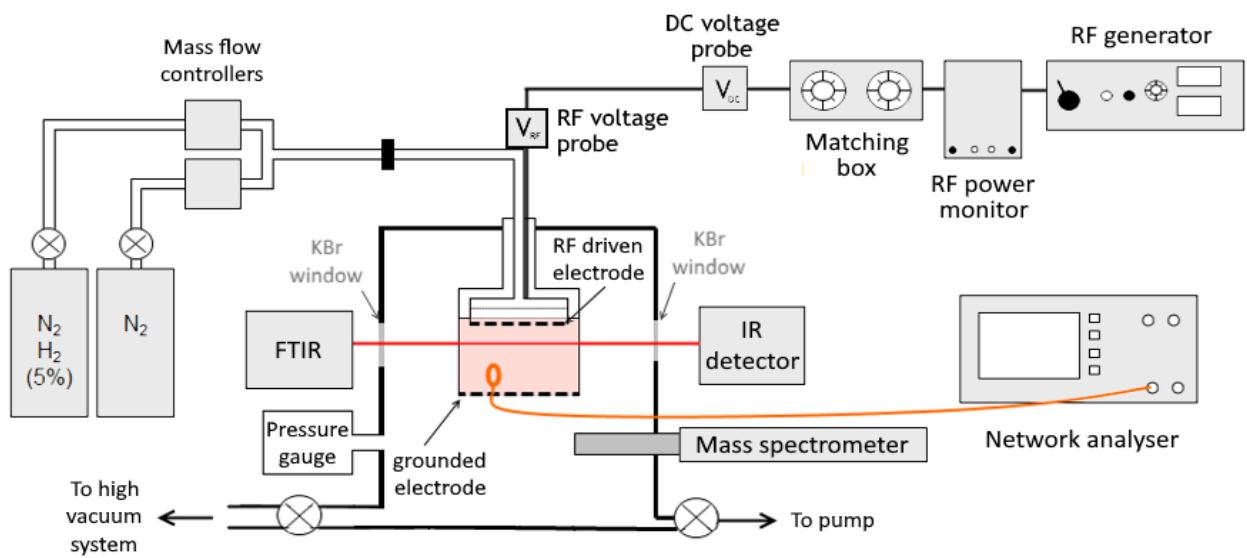

**Figure 1.** Experimental set up.

The experiment is a radiofrequency capacitively coupled plasma discharge (RF CCP) set up presented in figure 1 [22,40]. It is a wide stainless steel cylindrical chamber, with a diameter of 30 cm and a height of 40 cm. Two KBr windows are disposed for IR absorption measurements. In the middle of the chamber, the RF voltage is applied to the upper electrode which is a stainless steel disk grid of 12.6 cm in diameter. Through this grid the gas is injected as a homogeneous flow. The plasma is confined in a grounded aluminium alloy cylindrical box surrounding the upper electrode. The bottom of the box at 3.4 cm from the upper electrode is also a grid to let the gas go through. Four 2 cm-large and 4 cm-high slits are pierced in the sides of the box for the instruments to access the plasma. There are covered by thin metallic grids when unused in order to maintain the plasma in the box.

Some parameters are varied for this study: the percentage of $H_2$, the total pressure and the injected RF power. Before each experiment the chamber is heated and pumped to high vacuum (down to $10^{-6}$ mbar) with a turbo-molecular pump in order to clean the chamber. At the beginning of each experiment, the selected gas mixture is injected in continuous flow. The gas flow stabilizes after ~3 min and only then the plasma is ignited.

High purity gases (Air Liquide N60 Alphagaz 2, purity > 99.999%) are used. The amount of $H_2$ from 0 to 5% is obtained by mixing pure $N_2$ and a 95% $N_2$ - 5% $H_2$ mixture. The mixing is obtained by two mass flow controllers (MKS 100 Standard Cubic Centimetre per Minute (sccm), full scale accuracy 1%) injecting gas with individual flow varying from 2 to 70 sccm, and giving a global gas flow from 10 to 70 sccm. Pressure, measured by a capacitance gauge (MKS baratron 100 mbar full scale accuracy 0.15%), can therefore be adjusted from 0.3 to 1 mbar. The RF power generator delivers an incident power of 5 to 30 W at 13.56 MHz (SAIREM GRP01KE – 100 W maximum power). A matching box is connected between the generator and the plasma for impedance adaptation.

*2.2 Electrical measurements*

The RF peak-to-peak voltage ($V_{RF,pp}$) is one of the key reference parameters used to describe the experimental conditions. It is measured by a high-voltage probe connected to the driven electrode by a stainless steel tube of 30 cm in length, which induces almost no potential drop. It is positioned under vacuum to avoid electrical breakdown. A DC voltage probe situated between the RF driven electrode and the matching box gives the self-bias potential $V_{DC}$ with a precision of 0.3V. The self-bias potential appears when both electrodes differ in size and when a coupling capacitor is present between the RF power supply and the electrode (the matching box in figure 1). In these conditions, the asymmetry of currents collected on the two electrodes creates the self-bias potential [41]. It is related to the electron density and the electron temperature.

RF power measurements are performed by a digital V-I probe (Vigilent Power Monitor Solayl) positioned between the generator and the matching box. It gives the transmitted power with an accuracy of 2%. The incident and reflected power and the RF current (± 0.01A) are measured. The transmitted power measured with plasma ON is subtracted by the transmitted power with plasma OFF for the same RF peak-to-peak voltage to obtain the power absorbed by the plasma.

*2.3 Electron density measured by a resonant cavity method*

One of the most relevant parameter of the plasma is the electron density. It is determined in various sets of parameters thanks to a resonant cavity method. The theory of the resonant cavity method is explained by Slater [42], Klein, Donovan, Dressel et al [43–45]. It has been applied to a RF CCP discharge by Haverlag et al [46] and by Dongen [47]. Previous works already used this method in our reactor, in $N_2$-$CH_4$ [40,48,49] and pure $N_2$ [38] plasmas. This technique has the advantage of being non-intrusive.

The metallic confining box is used as a microwave resonance cavity. Microwaves are emitted and measured by an antenna loop of 0.8 cm in diameter. The antenna is positioned perpendicularly to a box radius to enhance the transverse magnetic modes (TM). A second loop can be positioned symmetrically to measure the transmitted signal. Measurements done in transmission with two antennas and in reflection with only one antenna have been compared (see appendix A1). We found that reflection measurements are more precise, so this mode was chosen in the following. Microwaves from 1 to 4 GHz are delivered and detected with a network analyser (Rohde & Schwarz ZVL Accuracy of reflection measurements: < 0.4 dB for 0 dB to -15 dB and < 1dB for -15 dB to -25 dB).

Microwaves emitted in a metallic cylindrical box ($R_c$ = 6.9 cm, $L_c$ = 3.5 cm) theoretically lead to resonant $TM_{l,m,p}$ modes described by:

$$f_{lmp} = \frac{c}{2\pi\sqrt{\varepsilon_r}} \sqrt{\frac{\lambda_{lm}^2}{R_c^2} + \frac{p^2\pi^2}{L_c^2}} \qquad l, p = 0,1,2 \ldots \qquad m = 1,2 \ldots \tag{1}$$

where $\lambda_{lm}$ is the $m^{th}$ zero of the equation $J_l(\lambda) = 0$, where $J_l$ is the $l^{th}$ Bessel function. In air or vacuum, in the configuration used here, the theoretical resonant frequencies lower than 4 GHz are the transverse magnetic modes $TM_{010}$ (1.66 GHz), $TM_{110}$ (2.65 GHz) and $TM_{210}$ (3.55 GHz).

The dielectric constant $\varepsilon_r$ of the plasma is a function of the electron density. It induces a shift in the resonance frequency with and without plasma, which gives the relation: [46]

$$n_e = A \times \frac{8\pi^2 m_e \epsilon_0}{e^2} \times \frac{f^2}{f_0} \times (f - f_0) \tag{2}$$

where $n_e$, $m_e$ and $e$ are respectively the electron density, mass and charge, $f$ and $f_0$ the resonant frequencies respectively with and without plasma. Finally, $A$ is a factor depending on the cavity geometry and the chosen mode [38]. It was calculated for pure $N_2$ plasmas taking into account $n_e$ profiles in our conditions: 1.03 for $TM_{010}$, 1.09 for $TM_{110}$ and 1.14 for $TM_{210}$ [50].

Improvements have been done on the technique since [38,40,49]. Microwave resonance spectra directly acquired in the confining box show parasitic absorption. These absorptions are attributed to the gap between the RF driven electrode and the grounded box. We added a thin copper crown in between the RF driven electrode and the box sides in order to create a more ideal cylindrical cavity. Parasitic absorption peaks disappear with this method (see appendix A1). Figure 2 shows a scheme of the addition of the copper crown in the experiment, and the resonance shifts obtained for a $N_2$ $H_2$ (95-5) mixture at 0.9 mbar.

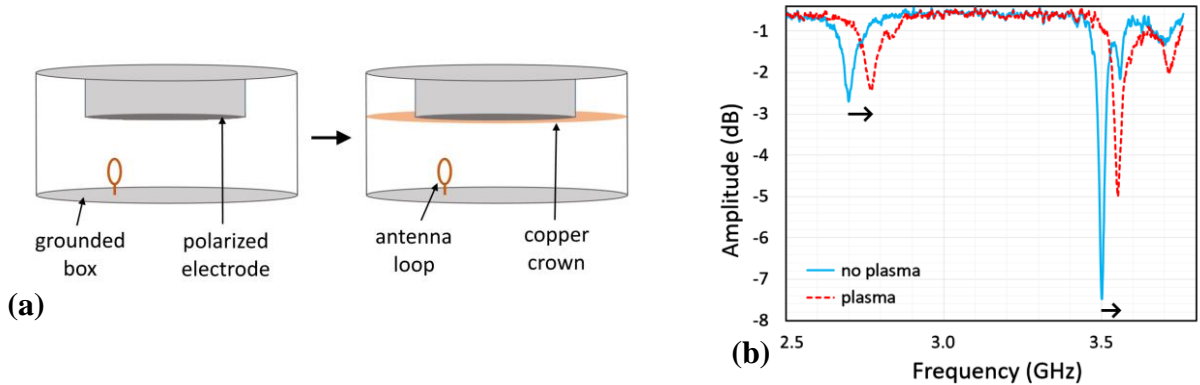

**Figure 2.** Electron density measurement. (a) Scheme of the installation of a copper crown between the RF driven electrode and the confining grounded box to improve the signal. (b) Spectra showing the resonant frequencies in the box, before and after plasma ignition (0.91 mbar – $N_2$-$H_2$ 5% - 13.5 W)

Two peaks are especially visible at 2.70 and 3.50 GHz, attributed to the modes $TM_{110}$ and $TM_{210}$. Measurements with both frequencies give similar results for the electron density, with differences of only 2%. Therefore, we then focused only on the $TM_{210}$ mode because it had a greater intensity. The resolution of frequency measurements is ~0.5 MHz, which gives a resolution of ~$6.10^7$ cm$^{-3}$ for electron density.

We observe a slow shift of the resonance when the plasma is turned OFF due to the box cooling, changing slightly its dimensions and therefore the resonance frequency. Consequently, $f_0$ is measured just after plasma extinction for all the experimental conditions. Each measurement was done at least 2 to 3 times on different days to ensure the repeatability. For measurements with identical $H_2$ amount, pressure and RF peak-to-peak voltage, the mean standard deviations are ~8% for the electron density, ~5% for the transmitted power and ~2% for the DC voltage.

## *2.4 Ammonia by IR transmission spectroscopy and neutral mass spectrometry*

### *2.4.1 Quantification of ammonia density by IR transmission spectroscopy*

Simple gaseous molecules have previously been studied in the experiment using IR spectroscopy. The plasma was ignited during hours and molecules accumulated in an external [51] or internal [52] cold trap. Thereafter, the content of the trap was released in the chamber for the IR analysis. The main objective was to obtain densities high enough for an accurate IR measurement. This method was tested for ammonia but without a good accuracy because ammonia adsorbs easily on metallic surfaces. Therefore, direct measurements during the plasma are necessary. Even if ammonia is homogeneously distributed in the chamber (see appendix A2), its density is low and the absorption length between the two KBr windows (50.8 cm) is not long enough to easily measure the absorption. The parameters of the FTIR spectrometer have to be finely adjusted to maximize the signal. We focused on a short wavelength range centred on the two most intense $NH_3$ IR absorption bands (between 850 and 1050 $cm^{-1}$).

Spectra are taken by a Fourier Transform IR spectrometer (FTIR – Nicolet 6700 from Thermo Fisher), with a MCT detector and a Michelson speed of 0.63 cm/s. 3000 to 6000 scans are accumulated, at resolutions between 1 and 4 $cm^{-1}$. A single final spectrum takes 2 to 3 hours to be obtained. The IR analysis is used for calibration of the MS (see part 2.4.2 and appendix A2).

Final spectra are analysed with a program developed by Klarenaar et al. [53]. It deduces the ammonia amount from the IR absorption spectrum, the experimental conditions (mainly total pressure) and the molecular data from the HITRAN database. The main uncertainty on the deduced value of ammonia density is the neutral gas temperature which requires a higher resolution to be deduced from the IR spectra. A Resistance Temperature Detector (RTD, PT 100) measures the RF driven electrode temperature. This temperature increases from 25 to 60°C during the 3-hours plasma duration. The gas temperature variations lead to 15-20% uncertainty on the ammonia density calculation. Figure 3 presents a typical IR absorption spectra recorded for a plasma of 3% of $H_2$ at 0.53 mbar.

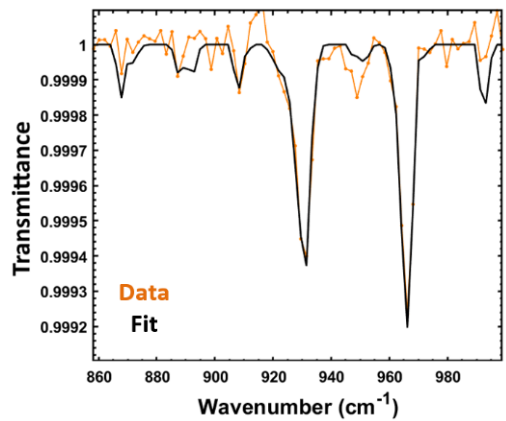

**Figure 3.** IR spectra of ammonia detected in the experiment ($N_2$-$H_2$ 3% - 0.53 mbar – 8.4 W) and fitted to deduce ammonia density. Spectra integrated over 1h45 at 4 $cm^{-1}$ resolution.

Eight spectra were acquired for different conditions: for amounts of H$_2$ of 1, 3 and 5% and at two different pressures, 0.53 mbar and 0.91 mbar.

*2.4.2 Calibration of the mass spectrometer with simultaneous IR spectroscopy measurements*

The ammonia amount is measured in the chamber by a quadrupole mass spectrometer (MS – Hiden Analytical, Electrostatic Quadrupole Plasma (EQP) series). The movable grounded sampling stick is positioned at a few centimetres from the confining box, with a sampling orifice of 100 µm. Ammonia is detected with the residual gas analyser mode (RGA), with an electron energy of 70 V and a filament intensity of 5 µA. The detector is a Secondary Electron Multiplier (SEM). Spectra were acquired with m/z ratio from 2 to 60, averaged on 3 to 10 scans, with a total acquisition time of 1-2 minutes, and a dynamic range of $10^6$ counts/s.

Ammonia fragmentation main ion peaks in RGA are at m/z 17 and 16. Traces of water are detected at m/z 18. We deduced from the peak at m/z 18 and the NIST database the minor contributions of water to the peaks at m/z 17 and 16 and removed it to obtain the contribution of ammonia (5-20% correction). Figure 4 presents mass spectra recorded for 3 gas mixture conditions.

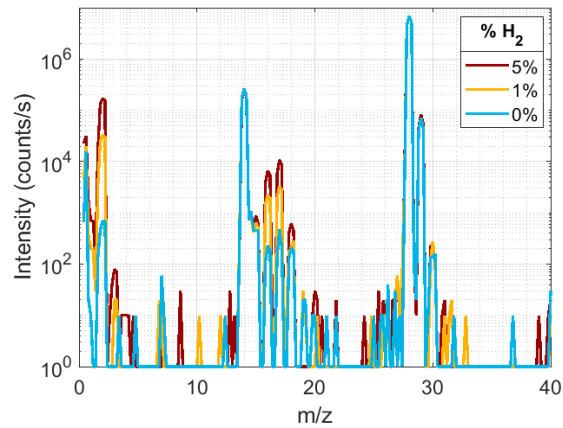

**Figure 4.** MS spectra during plasma (0.91 mbar, ~11 W) in N$_2$ and N$_2$-H$_2$ at different percentages of H$_2$. N$_2$ has fragments at m/z 7, 14, 15, 28, 29, 30. H$_2$ is visible at m/z 2. m/z 16 and 17 are mainly associated to ammonia. Traces of water are detected at m/z 18. The lower detection limit is at ~30 counts/s.

MS are highly sensitive to their electrical environment and measured intensities can drift. To avoid any changes in the global intensity due to charging effects by the plasma, intensities where divided by a reference peak in the spectra. We chose m/z 14, a dissociation peak of N$_2$ which is closer to the m/z 17 peak of NH$_3$ than m/z 28.

At the ignition of the discharge, close to the confining box, the ammonia amount roughly stabilizes in 2-3 minutes. However, ammonia tends to adsorb on the metallic walls of the chamber and the homogenization of ammonia in the chamber takes several tens of minutes depending on the surface state of the walls and the plasma parameters. Simultaneous measurements of ammonia by IR spectroscopy and MS spectrometry are therefore both started 30 min after the ignition of plasma, and integrated over 1.5 to 3 hours (depending on the parameters for the IR spectrum acquisition).

We performed two calibrations at two different pressures (see appendix A2). The precision is limited by the IR measurement (+/- 10%). The evolution of the calibration coefficient between NH$_3$ density and m/z ratio 17/14 is linear with pressure.

*2.4.3 Measurements of ammonia density by mass spectrometry*

Once the MS is calibrated at m/z 17 thanks to IR spectroscopy, it is useful to measure fast and local modifications of ammonia density in the chamber. The influence of $H_2$ percentage, RF plasma voltage and pressure are studied.

As ammonia tends to adsorb on the reactor walls, a great care was taken to have a clean wall surface state before each experiment. After a long exposure to plasma, walls are heated for several hours to eliminate residual ammonia. However, we observed that for short plasma durations (11 min), the use of a high vacuum pump between two experiments is sufficient. The MS took continuous measurements on a few selected masses (at m/z 14, 16, 17, 18, 28), starting 30 s before the ignition of the plasma and continuing during the 10 following minutes of the discharge. Ammonia intensity is roughly stable after a few minutes (see appendix A3). The intensity was integrated over the last few minutes where the signal is stable to get the final value for the deduction of the $NH_3$ density.

## *2.5 Positive ions measured by mass spectrometry*

The positive ion mode of the MS is used to detect positive ions formed in the discharge. A small hole is pierced in the confining box and the collecting head of the MS is positioned in front, in contact with the box to keep the plasma confined inside. A picture and a scheme of the configuration are given in the appendix A3. Consequently, the MS measures ions present next to the walls inside the confining box, at ~1 cm from the upper RF driven electrode. The collecting hole of the mass spectrometer and the confining box are both at the ground. Figure 5 shows typical MS spectra taken in ion mode in $N_2$ and $N_2$-$H_2$ plasmas.

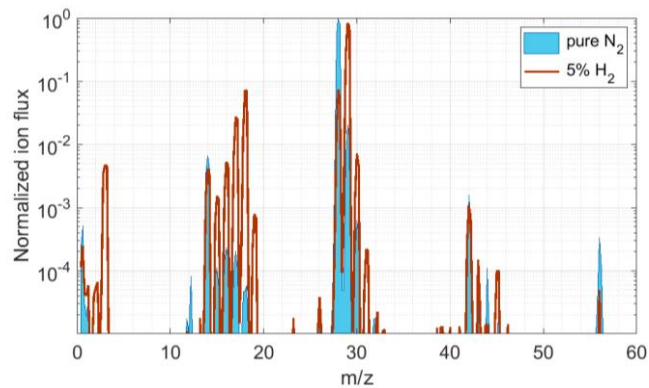

**Figure 5.** MS normalized spectra of ions in $N_2$ and $N_2$ – 5% $H_2$ discharges, at 0.86 mbar (55 sccm) and for an absorbed power of ~11 W (406 $V_{RF,pp}$)

| m/z | corresponding ion | m/z | corresponding ion |
|---|---|---|---|
| 2 | $H_2^+$ | 28 | $N_2^+$ |
| 3 | $H_3^+$ | 29 | $N_2H^+$ / isotope $N_2^+$ |
| 14 | $N^+$ | 30 | $N_2H_2^+$ / isotope $N_2H^+$ |
| 15 | $NH^+$ / isotope $N^+$ | 42 | $N_3^+$ |
| 16 | $NH_2^+$ | 43 | $N_3H^+$ |
| 17 | $NH_3^+$ | 44 | $N_3H_2^+$ |
| 18 | $NH_4^+$ | 45 | $N_3H_3^+$ |
| 19 | isotope $NH_4^+$ / $H_3O^+$ | 56 | $N_4^+$ |

**Table 1.** Selected ions for the study

MS acquisition parameters tuned for each mass are the same for m/z from 14 to 56, but are different for m/z 2 and 3. Therefore, for optimized measurements for all m/z, we used two different sets of parameters: one optimized for m/z 2 and 3, and one optimized for the others m/z. At such low power conditions, only ions with a charge z=1 are expected to be detected. Selected ion peaks at given m/z values (see figure 5 and table 1) are measured in continuous acquisition by the MS in ion mode from the ignition of the plasma and until stabilization of the signal. The intensity of each ion is then averaged over few minutes after stabilization of the signal (from ~8 to 10 min, see appendix A3).

MS does not give reliable absolute values for ions. Indeed, the intensities measured depend strongly on MS selected parameters and sometimes vary during the acquisition due to charging effects. Therefore, we provide only relative intensity values. In the following, the ion intensities (or fluxes) are normalized to the total intensity (or flux) measured for the ions, which is approximately equal to the sum of the intensities measured for m/z 28, 29, 17, 18 and 14 (see figure 5).

The $^{15}N$ isotope of nitrogen is usually present in proportions equal to 0.36% of nitrogen atoms, and can therefore be detected by the MS, especially in pure $N_2$ at m/z 15 ($^{15}N^+$), 29 ($^{14}N^{15}N^+$) and 30 ($^{15}N^{15}N^+$) (see table 1). In $N_2$-$H_2$ discharges, it can also influence intensity measurements at m/z 19 ($^{15}NH_4^+$) and 30 ($^{14}N^{15}NH^+$). Isotopes are therefore taken into account to study the intensities of other ions at m/z 15, 19, 29 and 30. Water has a strong proton affinity and $H_2O$ adsorbed in small quantities on the walls of the reactor easily forms $H_3O^+$, which contributes at m/z 19. As $H_2O^+$ (m/z 18) should be present in far lower quantities than $H_3O^+$, the strong signal at m/z 18 can be entirely attributed to $NH_4^+$. Impurities from air should be seen at m/z 32 ($O_2^+$) and m/z 44 ($CO_2^+$) in both pure $N_2$ and $N_2$-$H_2$ plasmas. Figure 5 confirms that they can be neglected in this study.

## *2.6 Mass dependence of the mass spectrometer transmittance*

In both neutral and ion modes, intensities measured by the MS at different m/z cannot be directly compared as the ionization, the flight in the MS through the quadrupole and the detection depend on the mass and the intrinsic properties of the species. A transmittance curve has been obtained for the MS used in this work. Such curves can be found in the literature: for EQP MS from Hiden equipped with SEM [54–56] or for other MS [35,57,58]. They use the method described below, always using several gases among $H_2$, He, Ne, $N_2$, $O_2$, Ar, Kr and Xe. The curves obtained in the literature vary from one another but in all cases, the transmittance is mainly decreasing with m/z. We show here that the transmittance depends strongly on the parameters of the MS. In particular, we obtained very different transmittances for the MS parameters tuned at m/z 2 ($H_2$) or at m/z 28 ($N_2$) (see appendix A4). This shows the necessity to measure the MS transmittance with the parameters used for the acquisition of the experimental spectra.

### *2.6.1 Calibration for neutrals*

Different gases (or gas mixtures in well-known proportions) were injected in the chamber. The transmittance of the MS at the mass of the studied gas was deduced from the comparison of the intensity measured by the MS and its partial pressure in the chamber. As the temperature is the same for all neutrals, the density of an atom or molecule is directly linked to its partial pressure. The transmittance as a function of mass is obtained once the measurement is done with different gases at various masses. For better accuracy, several measurements were performed for a same gas at different pressures. Then, the slope of the intensity as function of the partial pressure was used to deduce the transmittance.

More precisely, the transmittance $T_{tot}$ through the total system for neutrals is defined by the following equation:

$$I_{MS}(m_X) = P_{X,0} \times \frac{1}{K_{iso,X}} \times T_{tot}(m_X) \qquad (3)$$

Here, $I_{MS}(m_X)$ (in counts/s) is the intensity measured by the MS at the mass $m_X$ of the main isotope of the molecule/atom X present in the chamber. $P_{X,0}$ is the partial pressure (in Pa) of X in the chamber. It is known in the case of the calibration, but it is the value we want to obtain in a usual experiment. $K_{iso,X}$ is a non-dimensional coefficient that takes into account the isotopes of the molecule/atom X, to compensate the fact that only the intensity at the mass of the main isotope is used (e.g. in the case of Krypton, $K_{iso,Kr} = 1.767$). Isotopes abundances are measured and are congruent with values found on the NIST database.

The total transmittance ($T_{tot}$, in counts.s$^{-1}$.Pa$^{-1}$) can be written as the product of three components: the transmittance through the 100 μm aperture at the entrance of the MS ($T_{ap}$, no unit), the transmittance through the ionization chamber ($T_{ioni,X}$, in m$^{-2}$.s$^{-1}$.Pa$^{-1}$) and the transmittance through the main part of the MS (i.e. the lenses, the quadrupole and the detector), $T_{MS}$ (in counts.m$^2$).

$$T_{tot}(m_X) = T_{ap}(m_X) \times T_{ioni,X} \times T_{MS}(m_X) \quad (4)$$

with: $T_{ap}(m_X) = \frac{P_{X,MS}}{P_{X,0}}$, $T_{ioni,X} = \frac{j_{X^+,MS}}{P_{X,MS}}$ and $T_{MS}(m_X) = \frac{I_{MS}}{j_{X^+,MS}} \times K_{iso,X}$

where $P_{X,0}$ and $P_{X,MS}$ are respectively the partial pressures of the molecule/atom X in the chamber and in the MS (in Pa), and $j_{X^+,MS}$ is the ion flux inside the MS (in m$^{-2}$.s$^{-1}$).

$T_{tot}(m)$ is obtained from the calibration with gases presented above. It enables to deduce the partial pressure of the neutral species measured by the MS using equation (3). Nevertheless, concerning the measurements of positive ions by the MS, the transmittance to use is $T_{MS}(m)$ (see next section). Therefore, we estimated the expressions of $T_{ap}(m_X)$ and $T_{ioni,X}$ with a theoretical work to deduce $T_{MS}(m)$ from $T_{tot}(m)$ (see details in appendix A4).

Figure 6 shows the transmittances $T_{tot}(m)$ and $T_{MS}(m)$ obtained with our MS, normalized at m/z 20. The absolute transmittance $T_{tot}(m)$ and its error bars are given in appendix A4. The curves are well-fitted by a log-normal law:

$$f(x; G, \mu, s) = \frac{G}{x \cdot s \cdot \sqrt{2\pi}} \cdot \exp\left(-\frac{(\ln(x) - \mu)^2}{2s^2}\right) \quad (5)$$

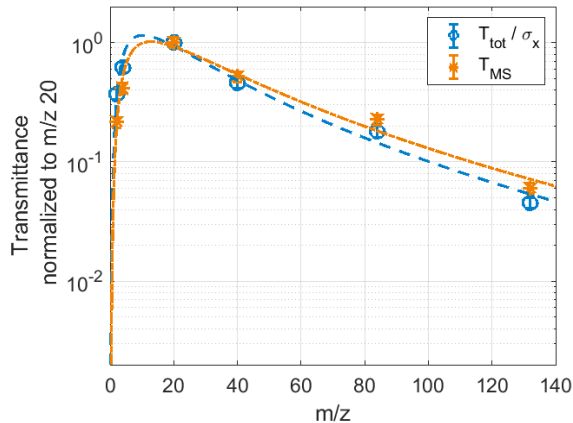

**Figure 6.** Transmittance obtained for a set of parameters tuned on m/z 28. Data points correspond to the following gases at their respective m/z: H$_2$ (2), He (4), Ne (20), Ar (40), Kr (84) and Xe (132). Comparison of the normalized transmittances $T_{tot}/\sigma_X$ and $T_{MS}$, fitted by log-normal laws with the following parameters: G$_{tot}$ = 52, G$_{MS}$ = 55, μ$_{tot}$ = 3.4, μ$_{MS}$ = 3.6 and s$_{tot}$ = s$_{MS}$ = 1.

These transmittances have been obtained for a given set of parameters of the MS, corresponding to a tune that optimizes m/z 28. The transmittance changes when the MS parameters are changed. In particular, it is very different for parameters tuned on m/z 2 (see appendix A4). In both cases, calibration curves have a steep slope at low m/z. To avoid any possible large errors, we do not compare intensities at m/z 2 and 3 with the higher m/z.

*2.6.2 Calibration for positive ions*

In the case of ions, there is no ionization source, but the addition of a high voltage just after the entrance of the MS (named the 'extractor') and an associated lens (named 'lens 1') to focus the incoming ions into the MS. It is far more complex to directly obtain the transmittance for ions as it is not usually easy to quantify ion densities or fluxes. Therefore, transmittances for ions are generally deduced from transmittances obtained with neutrals [35,56,59].

Ion transmittance through the lenses (except lens 1), the quadrupole and the detector is the same as in RGA mode ($T_{MS}$). As we work with ion fluxes, no change is expected from the passage through the 100 µm aperture [60]. The only part to study is the travel of ions through the extractor and lens 1: do they induce a mass-dependent ion transmittance? To address this issue, we studied the passage of ions of different masses through the extractor and lens 1. Details are given in appendix A5. We concluded that for absolute values of the extractor below 60 V and for ions with similar energy spectra, lens 1 focuses similarly all the ions (< 20-30% error), except for the low-mass ions (below m/z ~ 4). Therefore, we can expect that in these conditions, the group formed by the extractor and lens 1 has a transmittance independent on mass.

In conclusion, to compare ions measured on a same mass spectrum, with similar energies, we can use the above transmittance $T_{MS}$, obtained in RGA mode. It is valid for ions with m/z larger than ~5-10 u, and with a potential lower than ~60 V on the extractor. For an ion $X^+$ present in the chamber, leading to an intensity of $I_{MS}(m_X)$ detected by the MS, the relative ion flux in the chamber can be deduced by:

$$\frac{j_{X^+,0}}{j_{tot^+,0}} = \frac{j_{X^+,MS}}{j_{tot^+,MS}} = \frac{I_{MS}(m_X)}{T_{MS}(m_X)} \times \frac{1}{\sum_i \frac{I_{MS}(m_i)}{T_{MS}(m_i)}} \tag{6}$$

with $\Sigma_i$ being a sum on all the ions detected in the MS spectra.

The ion flux then computed is the ion flux at the entrance of the MS. A sheath forms on the MS collector head, and accelerates the ions towards the surface. Therefore, the ion flux measured is not equal to the ion flux in the bulk of the plasma discharge. Sode et al. [57] developed a simple model to take the sheath effect into account, to deduce the ion flux in the bulk, as well as the ion densities from the electron density. In the case of our RF CCP discharge, the 2D ion distribution is studied by a more complex model, detailed in paper II.

## 3. Results

Essentially three experimental parameters are varied to test their effect on the measurements. The $H_2$ percentage in the $N_2$-$H_2$ gas mixture varies from 0 to 5%. Pressure is varied from 0.33 to 0.91 mbar by changing the gas flow (from 10 to 70 sccm). RF peak-to-peak voltage is varied between 254 and 483 V.

### *3.1 Electrical measurements and electron density*

In the following figures, error bars take into account the accuracy and the measurements repeatability. Measurements are done at several selected peak-to-peak RF voltages ($V_{RF,pp}$: 254, 326, 367, 406 and 483 V). However, as mentioned in [38] the most relevant plasma parameter for the comparison between experiment and model is the RF power absorbed by the plasma ($W_{eff}$). The RF power measurements with the digital V-I probe (Solayl) (± 0.5 W) are much more accurate than measurements done in [38] (±2W). The measured absorbed power is given for all the conditions studied and plotted according to the RF peak-to-peak voltage in figure 7 (a).

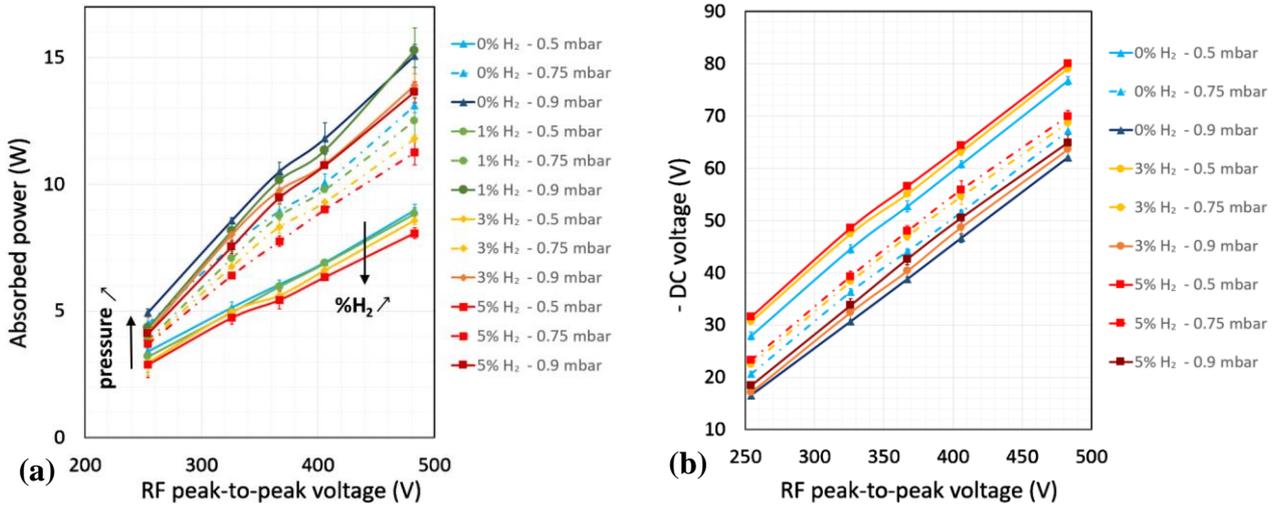

**Figure 7.** (a) RF power absorbed by the plasma and (b) self-bias potential $V_{DC}$ as function of $V_{RF,pp}$ in all the conditions studied here (various $V_{RF,pp}$, % of $H_2$ and pressure).

Absorbed power increases systematically with the RF peak-to-peak voltage. However, $H_2$ amount and pressure also have an effect on its value. At a given percentage of $H_2$ and a given $V_{RF,pp}$, the absorbed power increases strongly with pressure. At a given $V_{RF,pp}$ and pressure, the absorbed power decreases slightly with the amount of $H_2$.

The figure 7 (b) presents the self-bias potential ($V_{DC}$ < 0 in this configuration) for all the experimental conditions. These measurements are fundamental to validate the model described in paper II. The experiments show a linear dependence of $|V_{DC}|$ against $V_{RF,pp}$, with a constant slope of 0.2 in all conditions of pressure and hydrogen content. The variation of $|V_{DC}|$ with the discharge parameters relates to changes in the ion current [61]. In an asymmetric reactor, a higher ion current increases the asymmetry in the collection of charges at the electrodes, leading to a higher $|V_{DC}|$. At high $V_{RF,pp}$ the absorbed power increases, inducing an increase in the electron density (cf figure 8 (a)), the ion current and thus $|V_{DC}|$. On the other hand, an increase in the pressure leads to the reduction of the sheath thickness [61], implying a decrease in the ion current and $|V_{DC}|$, consistent with previous observations in pure $N_2$ [38].

Figure 8 (a) presents electron density results as a function of absorbed RF power. The experiments show that the electron density increases exponentially with $W_{eff}$. At a given $W_{eff}$, an increase in the pressure leads to lower electron densities, but the exponential behaviour is maintained. Similarly, figure 8 (b) shows that the electron density as a function of $V_{RF,pp}$ exhibits also an exponential evolution, but that it does not vary with pressure. For pure $N_2$ the $n_e$ values are globally in agreement with published values by Alves et al [38] and feature a comparable exponential increase. This exponential trend is specific to RF CCP discharges. While a regime similar to the positive column of a DC glow discharge happens at low pressure (<10-50 Pa) and low power, a different regime is observed in RF CCP discharges at higher pressures and powers, due to the efficient emission of secondary electrons from the walls, similarly to the negative glow of a DC discharge. The transition limit depends on the gas and the electrode material [62]. The transition regime has been observed in a helium discharge at 3 torr by Godyak et al [63] with a Langmuir probe, in He and Ar discharges at 1 torr by Greenberg et al [64] with microwave interferometry and in $O_2$ by Dongen [47] with a resonant cavity. It has been modelled by several works [65–67], and is taken into account in the model detailed in paper II.

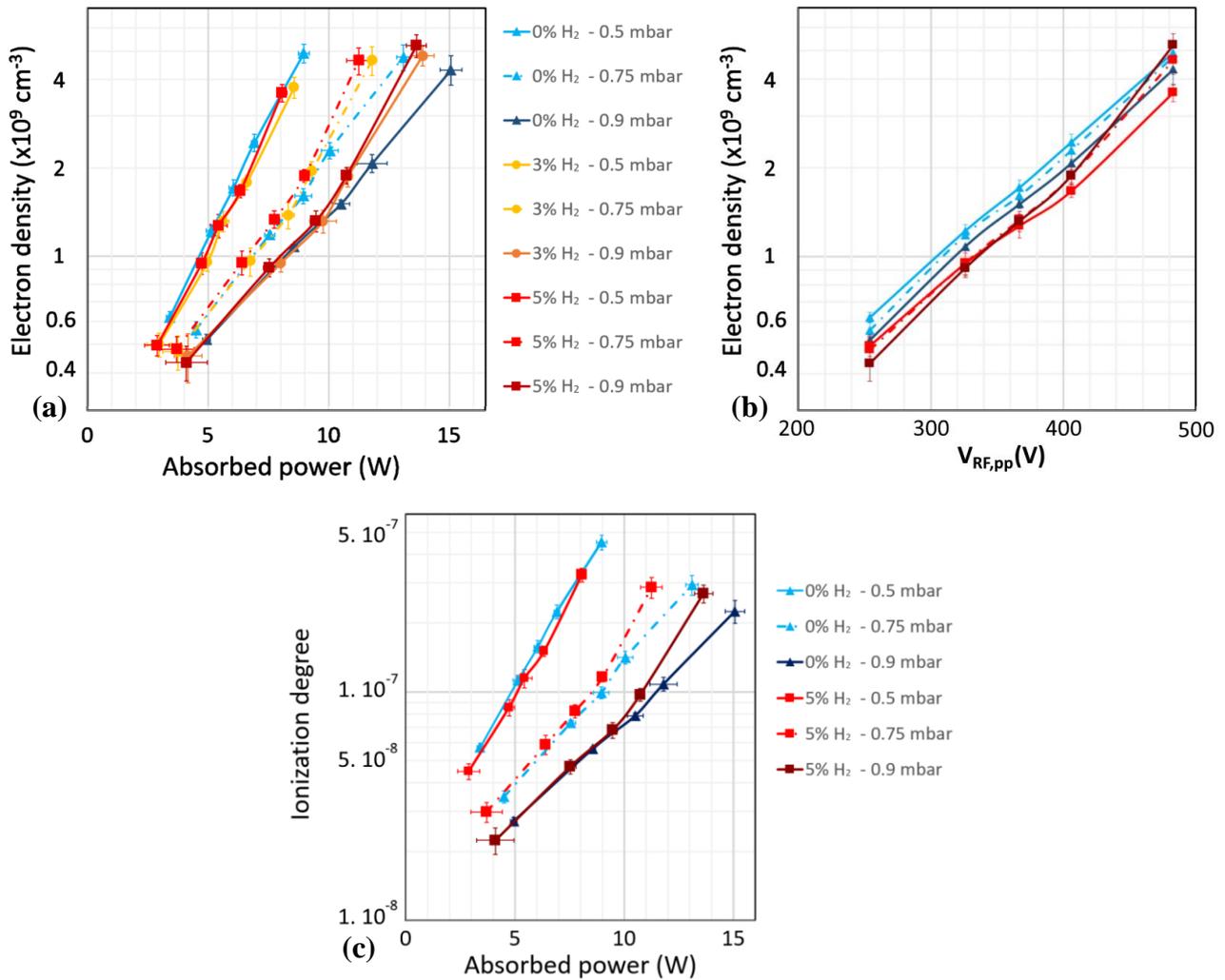

**Figure 8.** (a,b) Electron density and (c) ionization degree as function of absorbed power and $V_{RF,pp}$ for several experimental conditions.

The percentage of $H_2$ in the mixture does not have any effect on the electron density at a given pressure and absorbed power, except for the higher $W_{eff}$ cases. For high $W_{eff}$, despite large error bars, it seems that the increase of hydrogen content leads to an increase of the electron density. This point is discussed in paper II. The ionization degree is obtained from the electron density by dividing it by the total neutral density. To obtain the neutral density, the gas temperature was estimated as ~340 K +/- 10% from the measurements under similar conditions in $N_2$-$CH_4$ by Alcouffe et al. [40]. Figure 8 (c) shows that trends are the same as for electron density.

## *3.2 Ammonia density*

Ammonia is the most stable molecule formed in a $N_2$-$H_2$ plasma. The abundance of ammonia is a major constraint in the chemistry of the plasma (see paper II). A study of the formation of $NH_3$ depending on the different discharge conditions is shown in figure 9. For each condition $NH_3$ density is given in absolute values. An uncertainty of 20% is indicated on the graphs, evaluated from the uncertainty on the calibration coefficient and the MS measurements.

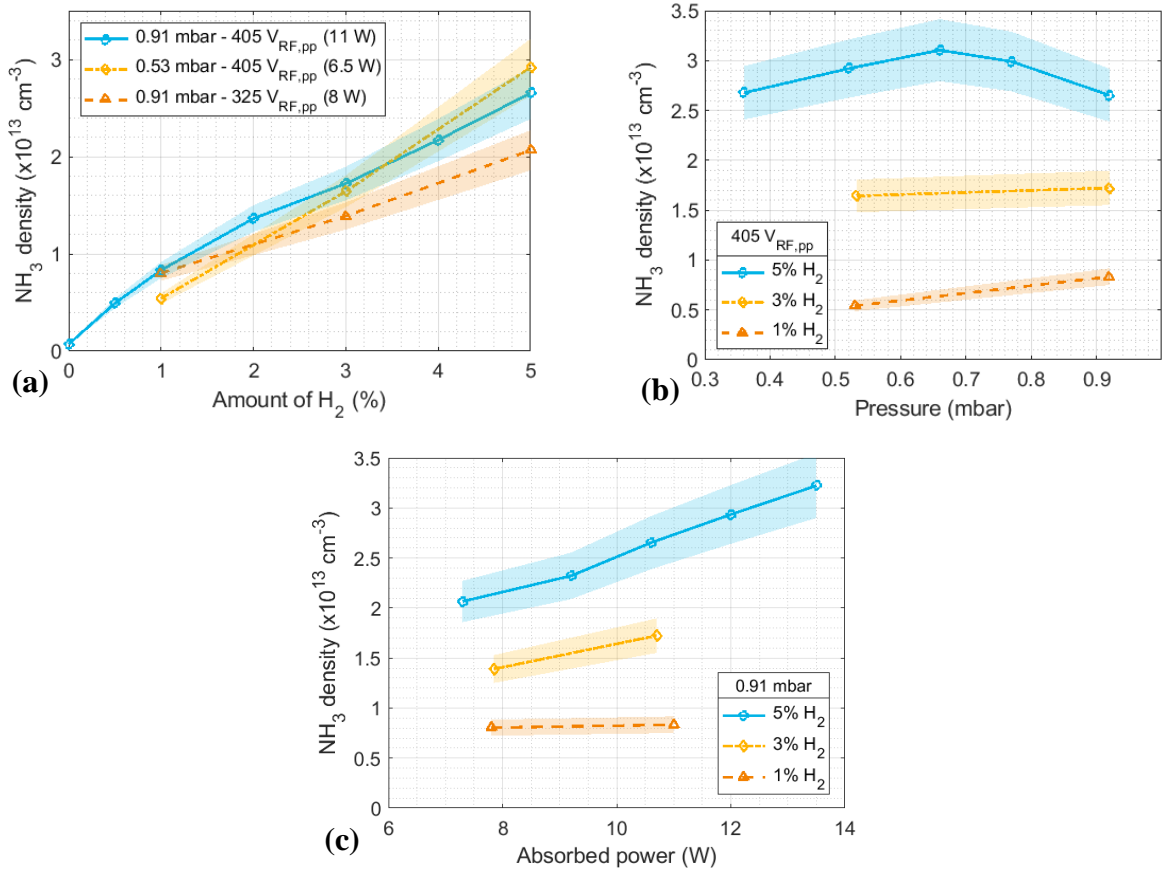

**Figure 9.** $NH_3$ amount as function of injected $H_2$ (a), pressure (b) and absorbed power (c). Note that for (a) and (b) the peak-to-peak RF voltage was kept constant, not the absorbed power. The relation between $V_{RF,pp}$ and the absorbed power is given in figure 7 (a).

Ammonia density increases linearly with the percentage of $H_2$ in $N_2$. It also increases slightly with the absorbed power, at $H_2$ concentrations above 1%. However, the $NH_3$ density variation with pressure from 0.35 to 0.9 mbar at a given $V_{pp,RF}$ stays within the 20% error bars. These large error bars are partly due to the fact that the wall surface state can vary slightly between two experiments (e.g. more N, H and/or $NH_3$ adsorbed on the walls) despite the cleaning protocol (see section 2.4.3). The small variation of $NH_3$ density with pressure can be linked to the small variation of the electron density with pressure, as showed in figure 8 (b). A nearly constant electron density is responsible for little changes in the flux of reactants towards the walls and the dissociation rate by electron impact, leading to a constant net production of ammonia. In addition, the small variation with pressure may also be due to the fact that the adsorption sites at the surfaces are saturated, which would strongly limit the production of ammonia that happens mainly on surfaces, as discussed in several works [30,31] and in paper II.

## *3.3 Positive ions*

The positive ions give important indications on the gas phase chemistry happening in the plasma and on nearby surfaces. They have been measured two to four times for each of the selected experimental conditions. Following figures present the average data points for each condition. Error bars are estimated from the reproducibility of the measurements and an uncertainty on the MS calibration estimated to +/- 20% (see section 2.6). These error bars take into account the instrumental (charging effects) and plasma effects (e.g. stabilization of the ion fluxes in the plasma as presented in appendix A3, and variations in the impurities quantity). The following figures present the evolution of ion fluxes with the parameters of the experiment ($H_2$ amount, pressure and absorbed power). These are the ion fluxes at the entrance of the MS. The 2D ion densities inside the reactor chamber are retrieved by the model presented in paper II.

Figure 10 shows the evolution of relative ion fluxes with the percentage of $H_2$ in the plasma. As explained in part 2.6, the analysis of m/z 2 and 3 is separated from the others. No calibration is applied to correct these measurements at low m/z, therefore no quantitative comparison can be done with the other m/z. For reproducibility (see part 2.5), the ion measurements are normalized to the sum of all ion fluxes (or to the sum of all intensities in the case of m/z 2 and 3).

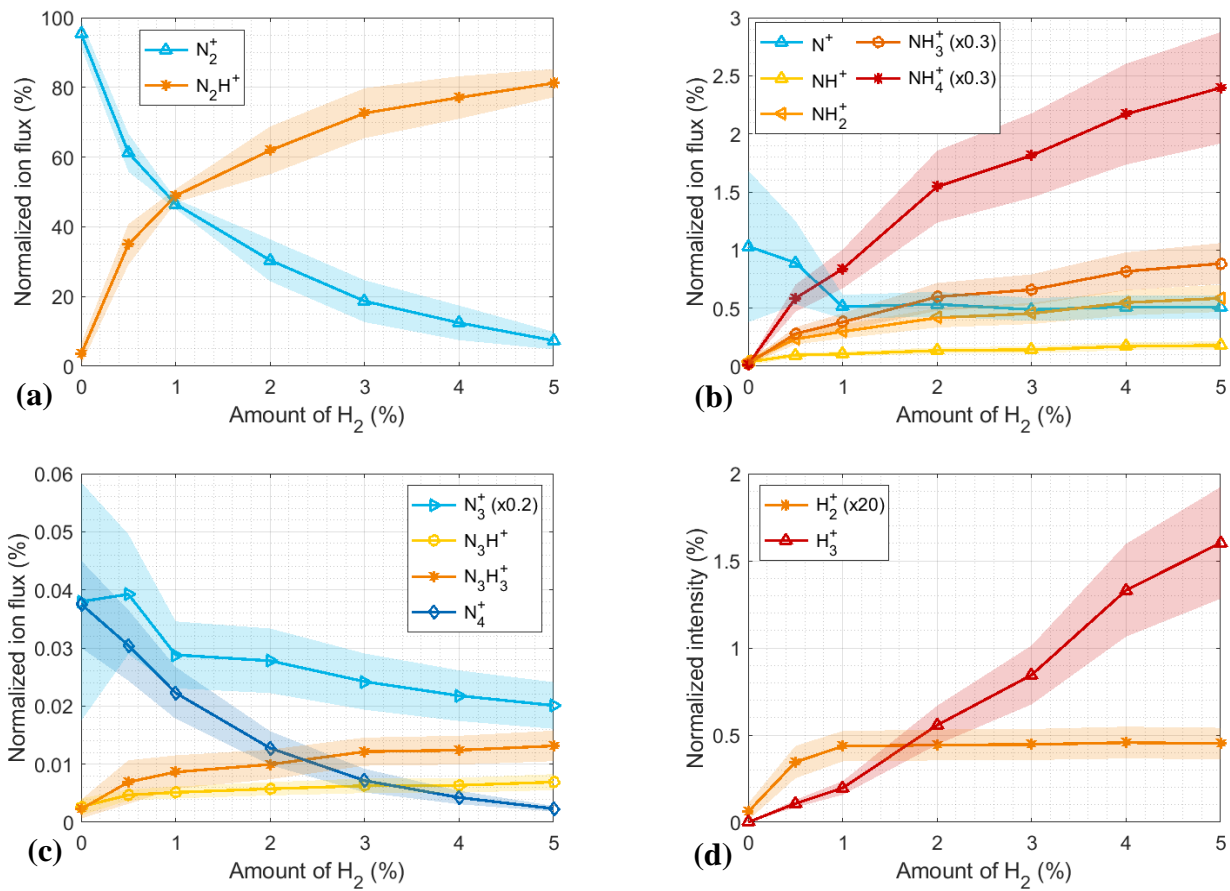

**Figure 10.** Normalized positive ion fluxes (a,b,c) or intensities (d) as function of the percentage of $H_2$ injected in the experiment. Conditions with 0.9 mbar (55 sccm) and ~10 W absorbed power (405 $V_{RF,pp}$). The curves for some species are scaled for representation purpose, the scaling coefficients are indicated in the legend.

The amount of $H_2$ injected in the plasma has a strong impact on the ions produced. In pure $N_2$ plasma, $N^+$, $N_2^+$, $N_3^+$ and $N_4^+$ are detected. These ions formed with only nitrogen decrease as soon as a small quantity of $H_2$ in injected in the plasma. $N_2^+$ and $N_4^+$ have nearly disappeared with 4-5% of $H_2$. $N^+$ and $N_3^+$ on one side, and $N_2^+$ and $N_4^+$ on the other side, have similar trends. Their formation and/or stabilization processes are linked.

On the other hand, new species containing hydrogen appear: nitrogen ions become protonated thanks to the injection of hydrogen. The $H_3^+$ ions density seems to be proportional to the injected $H_2$%, no saturation effect is seen for values of hydrogen inferior to 5%. On the contrary, $H_2^+$ stabilizes as soon as 1% of $H_2$ is injected in the experiment. $N_xH_y^+$ ions form in the presence of hydrogen, but their growth is not linear with the amount of $H_2$: their formation is limited by another variable than the quantity of hydrogen, contrarily to $H_3^+$. In particular, the $N_2H^+$ increase with the $H_2$ percentage slows down after 1% of $H_2$ injected. Its evolution is anti-correlated with the one of $N_2^+$: its formation is limited by the available quantity of ionized nitrogen. At 5% $H_2$ (at 0.9 mbar and ~10 W), $N_2^+$ decreases at 7.3% and the major ions after $N_2H^+$ (78%) are the ions $NH_4^+$ (10%) and $NH_3^+$ (3.7%) leading to the formation (or formed thanks to) ammonia. This is consistent with observations by [35] in a ICP discharge, where $N_2H^+$ is predicted to be dominant for percentages of $H_2$ lower than 60%, $NH_4^+$ and $NH_3^+$ being dominant for $H_2$ amounts above this value.

Ion populations are modified by the variation of pressure (induced by the variation of the gas flow from 10 to 70 sccm). Results are presented on figure 11.

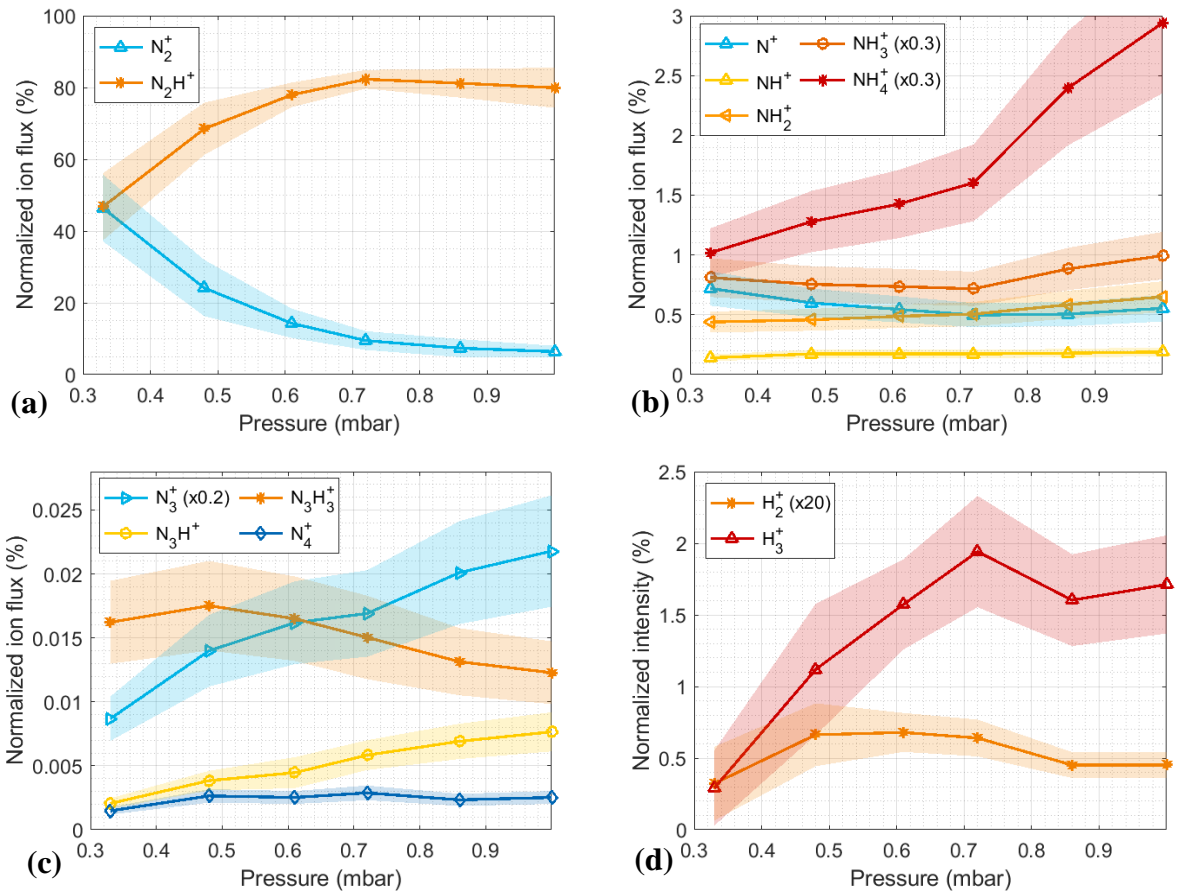

**Figure 11.** Normalized positive ion fluxes (a,b,c) or intensities (d) as function of pressure. Conditions with 5% $H_2$ and 6-11 W absorbed power (405 $V_{RF,pp}$). The curves for some species are scaled for representation purpose, the scaling coefficients are indicated in the legend. Note that the peak-to-peak RF voltage was kept constant, not the absorbed power. The relation between $V_{RF,pp}$ and the absorbed power is given in figure 7 (a).

The increase of the gas flow (and therefore the pressure) leads to an increase of nearly all the ions, except $N_2^+$ and $N^+$. Especially, at 0.33 mbar, $N_2^+$ and $N_2H^+$ have equal fluxes. It seems that at higher pressures, more complex and/or protonated ions are formed and/or transported in higher quantities to the walls of the confining box where the measurement is done. $N_2H^+$, $NH_4^+$, $H_3^+$, and in smaller proportions $N_3^+$, $N_4^+$ and $N_3H^+$ increase clearly with pressure, while $H_2^+$, $NH^+$, $NH_2^+$ and $NH_3^+$ stay rather constant. This can be easily explained by the fact that N$_2$H$^+$, NH$_4^+$, H$_3^+$, N$_3^+$, N$_4^+$ and N$_3$H$^+$ require a reaction between an ion (or radical) and a neutral gas phase molecule to form, whereas H$_2^+$, N$_2^+$, N$^+$, NH$^+$, NH$_2^+$ and NH$_3^+$ can be simply formed by direct ionization of neutrals H$_2$, N$_2$ and NH$_3$ (see table of reactions in paper II). With the increase of pressure, ion – neutral collisions increase and consequently ions formed from such collisions are formed in higher proportions compared to ions formed by direct ionization (see figure 13). Similar observations are discussed in [31].

Ions are also influenced by the power absorbed by the plasma (see figure 12).

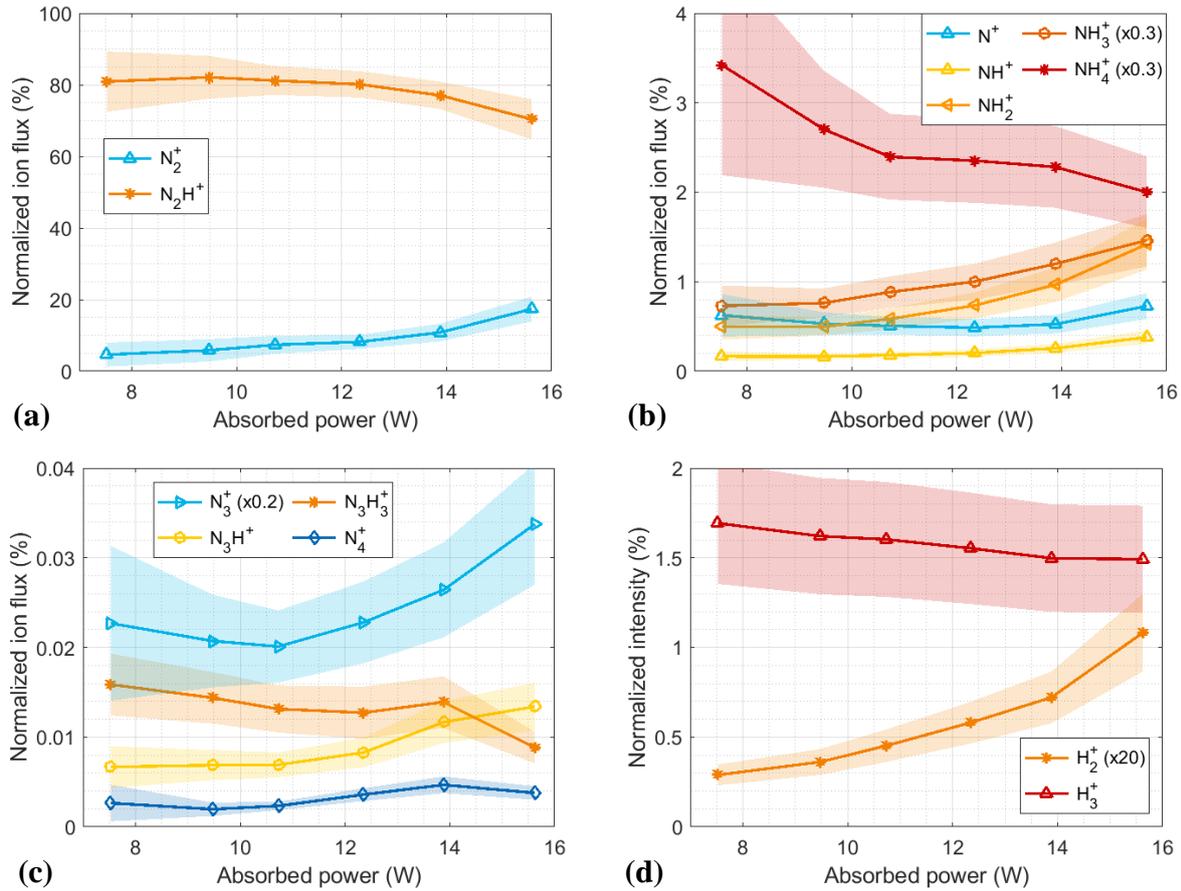

**Figure 12.** Normalized positive ion fluxes (a,b,c) or intensities (d) as function of the power aborbed by the plasma. Conditions with 5% H$_2$ and 0.9 mbar (55 sccm). The curves for some species are scaled for representation purpose, the scaling coefficients are indicated in the legend.

The increase of absorbed power moderately impacts the positive ion relative fluxes. However, the electron density increases, consequently the total ion density should increase similarly. A decrease of the relative intensities is observed for some species formed by ion (or radical) – neutral processes: of 10% for H$_3^+$, N$_2$H$^+$, N$_3$H$_3^+$ and 40% for NH$_4^+$. On the contrary, species formed by direct ionization increase: H$_2^+$, NH$^+$, NH$_2^+$, NH$_3^+$, N$_2^+$. Contrarily to the increase of pressure, an increase of power favours the formation of species by direct ionization. A small relative increase of N$_3^+$, N$_3$H$^+$ and N$_4^+$ can also be noted.

Figure 13, inspired from Carrasco et al. [17,31], summarizes the main reactions affecting the ion formation, which are also implemented in the model of paper II. The experimental results agree with the nature of the processes proposed here. We observe an increase in the density of the ions formed by direct ionization with increasing absorbed power, hence with increasing electron density that enhances direct ionization processes.

We also observe that ion-neutral and radical-neutral reactions favour the production of ions at high pressure. The model presented in paper II, after validation against the experimental results for the RF CCP discharge presented here, aims to quantify the relative importance of each production / destruction mechanism for electrons and heavy-species, identifying the most relevant chemical paths leading to the formation of ions and radicals, including the catalytic formation of ammonia at the discharge wall.

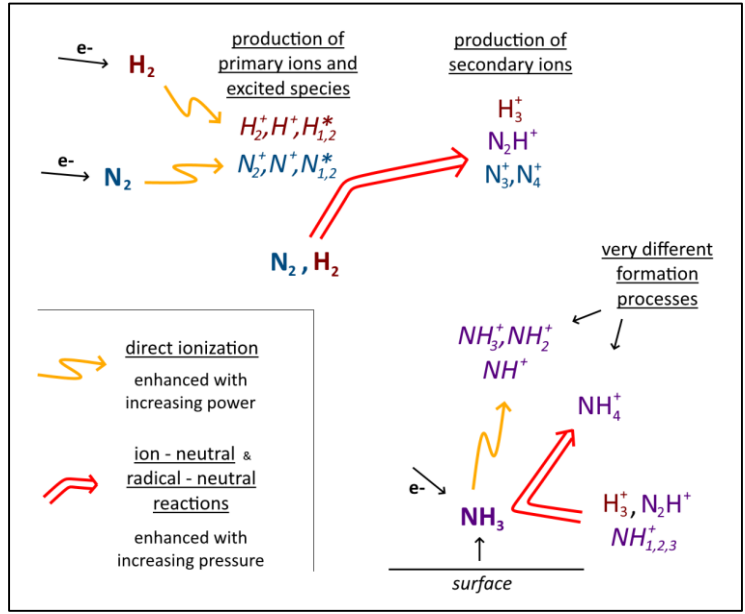

**Figure 13.** How coupled power and pressure affect ion formation: the main reactions.

## 4. Conclusion

In this work, a RF CCP discharge in nitrogen with small amounts of hydrogen (up to 5%), at low coupled power (3 to 13 W) and low pressure (0.3 to 1 mbar) is experimentally characterized. Electrical parameters, electron density, ammonia density and positive ions are measured in various different plasma conditions. One of the objectives is to form a complete dataset for the implementation of hydrogen-linked processes in the model of pure $N_2$ RF CPP discharge described in [38]. The model is described in paper II and the reader should refer to it for more details.

A complete study of the MS transmittance has been performed to deduce relative ion fluxes from MS measurements. The main conclusion is that the transmittance depends strongly on the set of MS parameters. In particular, sets of parameters obtained for a tuning on m/z 2 ($H_2$) lead to a MS transmittance very different than if tuned on m/z 28 ($N_2$).

The addition of hydrogen in an initially pure $N_2$ discharge induces some electrical changes. A 10-15% decrease of the coupled power for a same peak-to-peak RF potential is observed. However, the electron density stays rather constant with the $H_2$ amount, except for the higher pressure and power cases, where it could increase up to 25% with the addition of 5% hydrogen.

Concerning the molecular plasma species, the addition of $H_2$ has the expected effect to hydrogenate nitrogen ions, and lead to the formation of $NH_3$. $N_2H^+$ is the major ion in the discharge for $H_2$ amounts above 1% (~78% for 5% injected $H_2$), while $NH_3^+$ and $NH_4^+$ are the following dominant protonated ions (at respectively ~10% and ~3.5% for 5% injected $H_2$). The improvements to the model in pure $N_2$ thus focus on the addition of new protonated species and reactions using them as reactants or products. A particular attention is given to ammonia and surface processes (see paper II).

The variation of pressure from 0.3 to 1 mbar and coupled power from 3 to 13 W also leads to changes in the plasma behaviour. An exponential increase of the electron density is observed with the increase of coupled power. It is due to the production of secondary electrons on surfaces (see paper II). On the opposite, electron density decreases with the increase of pressure.

Ammonia quantity always increases similarly to $NH_2^+$ and $NH_3^+$, which suggests a strong dependence between these species, $NH_2^+$ and $NH_3^+$ being easily formed by direct ionization of $NH_3$. Ammonia absolute density stays constant with a pressure variation. Therefore, it suggests that ammonia formation is not governed only by bulk processes but mainly by surface processes. This point is discussed further in paper II. On the other hand, ions are formed in the bulk and disappear at the contact of surfaces. Their quantities are more sensitive to pressure variations. In particular, $NH_4^+$, $N_2H^+$, $H_3^+$, $N_3^+$, $N_4^+$ and $N_3H^+$, which require ion-neutral gas phase reactions to form, increase at higher pressure compared to other ions formed by direct ionization (especially $N_2^+$ and $N^+$). The increase of coupled power seems to have an opposite effect by enhancing the formation of ions by direct ionization, with the increase of $N_2^+$, $NH^+$, $NH_2^+$, $NH_3^+$, $H_2^+$ and the decrease of $NH_4^+$, $N_2H^+$ and $N_2H_2^+$.

On the present work we analysed new plasma conditions, different from the studies by [16,17,31,35]. Experiments were performed with only a few percent of $H_2$ (< 5% here, compared to > 40% in the studies cited above), at lower powers (~10 W compared to > 50 W) and higher pressures (~1 mbar compared to < 0.08 mbar). All these changes lead to a lower relative production of ammonia (0.15% compared to > 3%). Paper II discusses further the experimental observations by comparison with model results.

The conditions used here are very similar to plasmas in planetary atmospheres. Ionization degree is varied from $2.10^{-8}$ to $5.10^{-7}$ according to the studied parameters. These values are typically similar to ionization degrees in planetary atmospheres. In particular, it corresponds to the altitude range of 950 – 1100 km on Titan, where organic aerosols form in the $N_2$-$CH_4$-$H_2$ plasma [68]. This validates the idea of an experimental simulation of Titan's ionosphere with a RF CCP discharge [22]. The comprehension of $N_2$-$H_2$ plasmas is a first step in the analysis of $N_2$-$CH_4$-$H_2$ discharges that are chemically highly complex. In particular, [69] shows that $N_2$-$H_2$ plasma species are suspected to erode organic particles also present in the ionosphere of Titan. The understanding of $N_2$-$H_2$ laboratory plasmas can be very useful to understand the ion chemistry in the ionospheres of planets.

## Acknowledgements

N. Carrasco acknowledges the financial support of the European Research Council (ERC Starting Grant PRIMCHEM, Grant agreement no. 636829).

A. Chatain acknowledges ENS Paris-Saclay Doctoral Program. A. Chatain is grateful to Gilles Cartry and Thomas Gautier for fruitful discussions on the MS calibration.

L.L. Alves acknowledges the financial support of the Portuguese Foundation for Science and Technology (FCT) through the project UID/FIS/50010/2019 and grant SFRH/BSAB/150267/2019.

L. Marques and M. J. Redondo acknowledge the financial support of the Portuguese Foundation for Science and Technology (FCT) in the framework of the Strategic Funding UIDB/04650/2019 and project UTAP-EXPL/NTec/0107/2017.

# APPENDIX A1. Improvements of the resonant cavity method

Figure A1-1 presents these resonant TM modes obtained without and with the crown. These spectra are obtained in the configuration with two antennas (transmission mode). Especially, the $TM_{110}$ and $TM_{210}$ are more clearly identified with the crown.

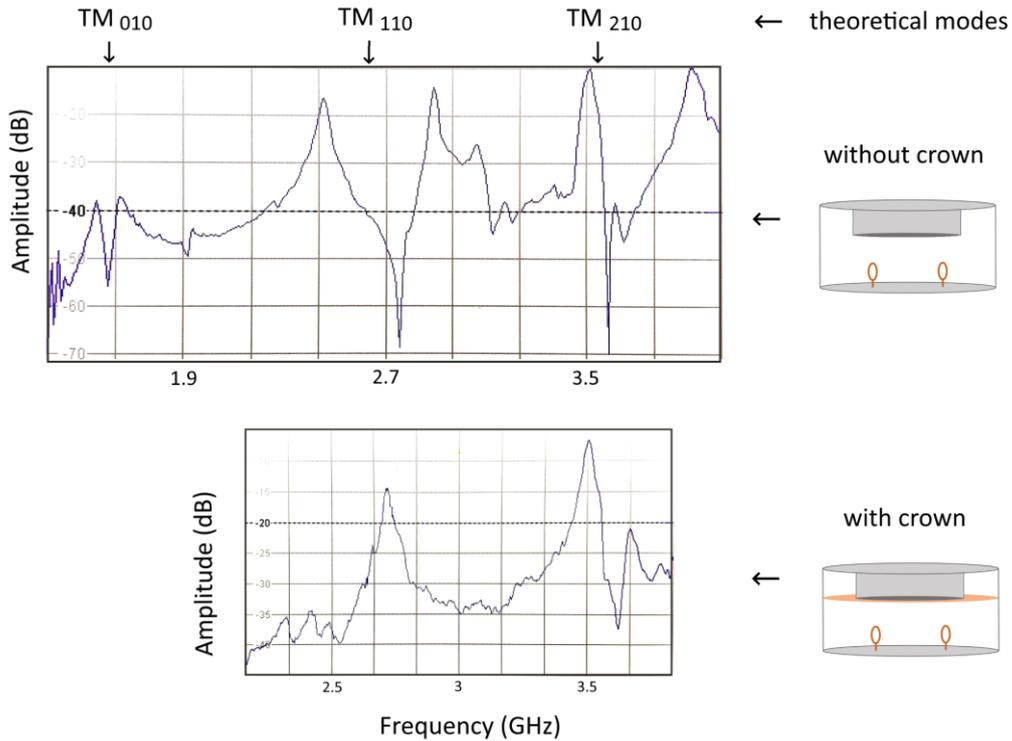

**Figure A1-1.** Spectra showing resonant TM modes in the box, without and with the copper crown.

The resonance is also detected by the refection on the emitting antenna. Spectra acquired in transmission (with two antennas) and in reflection (with one antenna) are presented figure A1-2. The signal in reflection has a more stable baseline and thinner peaks. As the signal for the $TM_{210}$ mode is the most intense, measurements of resonance shifts are done in reflection, and with this mode.

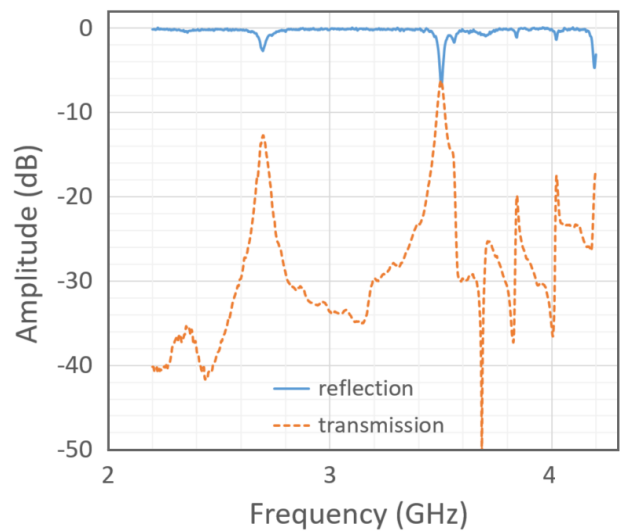

**Figure A1-2**. Resonances of the $TM_{110}$ $TM_{210}$ frequencies in the box, in reflection and in transmission.

# APPENDIX A2. Calibration of MS NH$_3$ measurements by IR spectroscopy

As mentioned above in the experimental section, IR absorption measurements are very long whereas mass spectrometry measurements are faster. But mass spectrometer gives only relative values.

Ammonia is created in the confined plasma at the center of the chamber; it diffuses through the grids of the metallic box, fills all the volume of the chamber and adsorbs on the chamber's walls. Depending on the plasma conditions, it can take several hours to reach the steady state of ammonia flux, and therefore a homogeneous ammonia spatial distribution in the chamber.

The spatial homogeneity and the reaching of the steady state of ammonia density in the chamber are studied with the mass spectrometer. Its collecting head can be radially moved from the metallic box to the wall of the chamber for measurements at different locations. Figure A2-1 shows that the spatial distribution of NH$_3$ is homogeneous. For calibration measurements, the IR spectrum is integrated over ~2h, and we similarly integrate the mass spectrum over the same acquisition period.

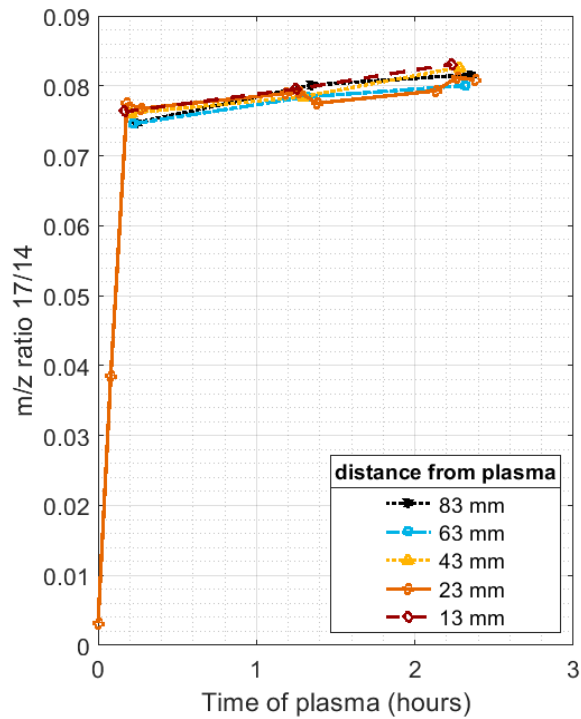

**Figure A2-1.** Variation of m/z ratio 17/14 (representative of ammonia) in time and distance from plasma (0.53 mbar – 3% H$_2$ – 8.4 W).

As NH$_3$ is homogeneous along a chamber diameter, the density can directly be deduced from the IR absorption. For a given experimental conditions, this density is compared with the ratio 17/14 of the mass spectrometer m/z intensities as presented figure A2-2.

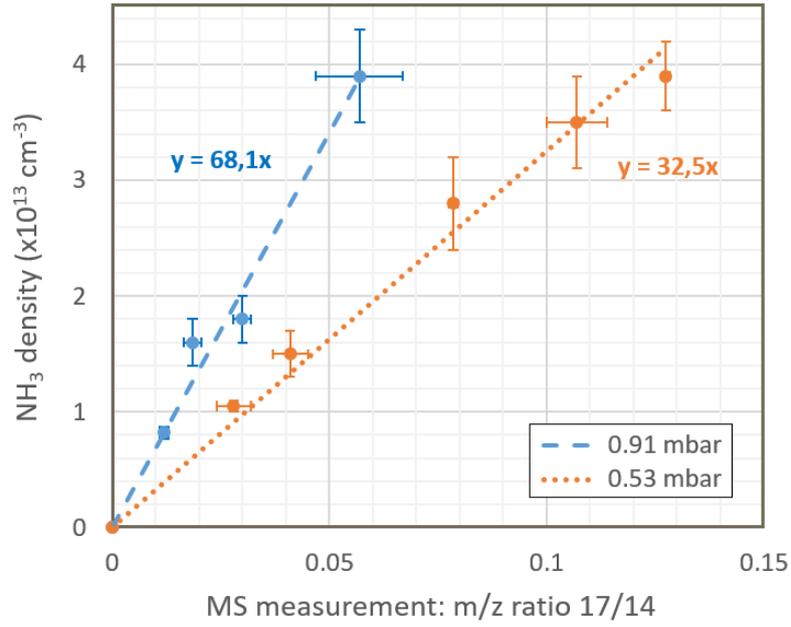

**Figure A2-2.** Calibration of MS measurements of $NH_3$ with IR data. Experiments done at 2 different pressures (0.91 and 0.53 mbar) in a $N_2$-$H_2$ plasma discharge, for a same RF peak-to-peak voltage of 480 V (at resp. 14 and 8.5 W).

The calibrations curves lead to the following calibration formula:

$$[NH_3] \approx \frac{I(17)}{I(14)} * 70 * P \; (.10^{13} \text{cm}^{-3}) \qquad [A2.1]$$

with $[NH_3]$ the average ammonia density in the chamber, $\frac{I(17)}{I(14)}$ the ratio of intensities measured at m/z 17 and m/z 14 by the mass spectrometer, corrected of any water contribution, and P the pressure in mbar.

# APPENDIX A3. Stabilization of MS measurements (neutrals and ions)

Neutrals and positive ions are measured continuously from the ignition of the plasma to the end of the experiment. Both measurements take some minutes to stabilize (see figure A3-1). This is mainly due to two effects: the stabilization of the species densities in the plasma and the stabilization of the MS measurement, which is likely to vary due to charging effects at the beginning of the acquisition. For the intensity measurements of neutrals and ions, the stabilization time is noted in all cases, and an average is done on the acquired signal only after the signal is stabilized.

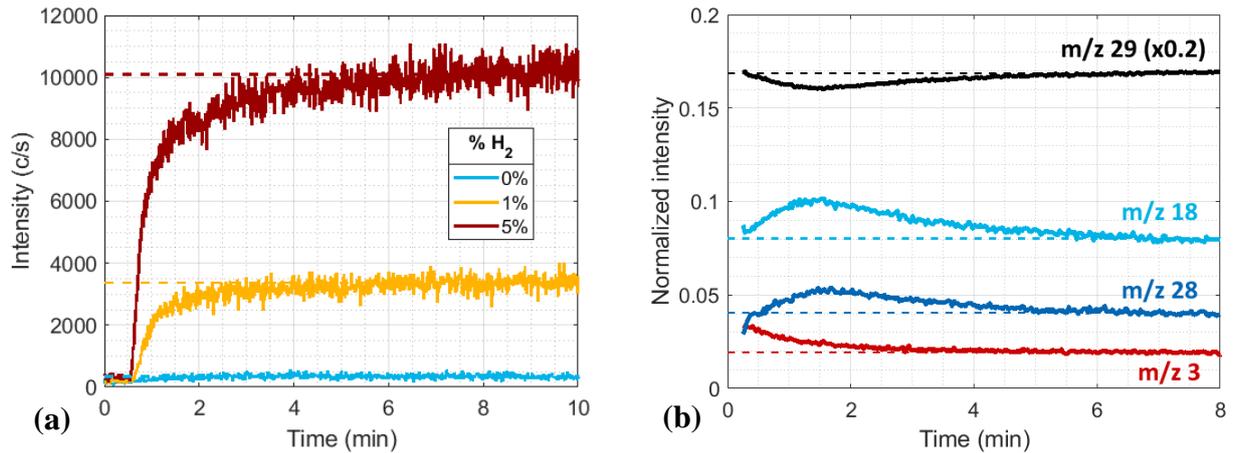

**Figure A3-1**. MS continuous acquisition in RGA and ion modes at the ignition of the plasma. (a) Intensity of m/z 17 (representative of $NH_3$ amount) at the ignition of $N_2$ and $N_2$-$H_2$ plasmas (0.53 mbar – 6.5 W). (b) Evolution of m/z 3 ($H_3^+$), 18 ($NH_4^+$), 28 ($N_2^+$) and 29 ($N_2H^+$) in time until stabilization (5% $H_2$ – 0.86 mbar – 10.5 W).

The acquisition of positive ions is possible thanks to a hole pierced in the plasma confining box. The MS collecting stick is approached in contact to the box, in front of the hole. Figure A3-2 shows a picture and a scheme of the configuration.

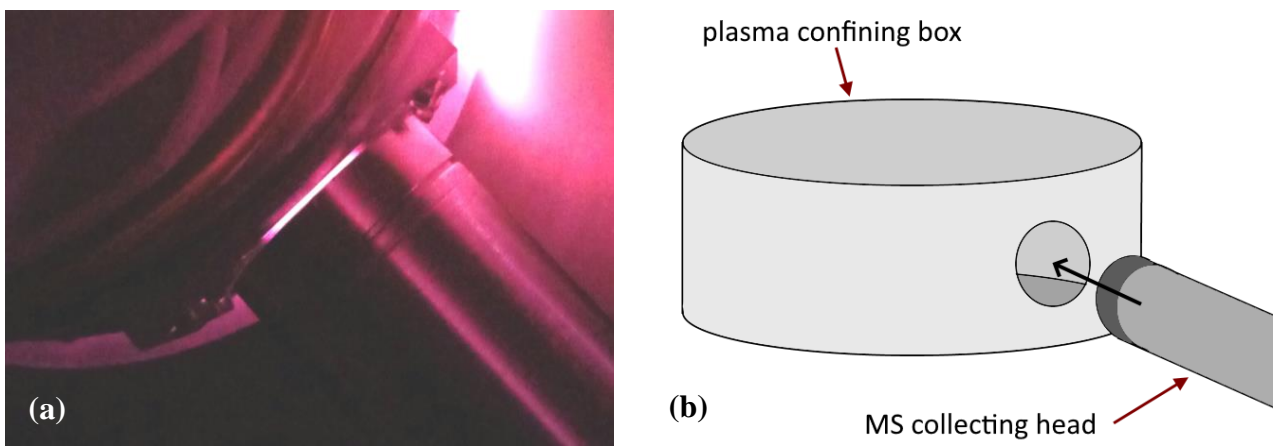

**Figure A3-2**. Configuration of the plasma confining box and the MS collecting head for the measurement of positive ions.

*Chatain et al (2020) PSST, accepted version*

# APPENDIX A4. Transmittance of the mass spectrometer as function of m/z

**The total transmittance and its error bars**

We computed the total transmittance through the MS ($T_{tot}$) as done in [57]:

$$\frac{T_{tot}(m_X)}{\sigma_X} = \frac{I_{MS}(m_X) \times K_{iso,X}}{\sigma_X \times P_X} \qquad (6)$$

with $\sigma_X$ the simple ionization cross section of the molecule/atom X, $P_X$ its partial pressure, $I_{MS}(m_X)$ the MS intensity at the mass $m_X$ of the main isotope of X and $K_{iso,X}$ an isotope corrective factor.

The final calibration curve is given in figure A4-1. It is well-fitted by a log-normal law:

$$f(x; G, \mu, s) = \frac{G}{x. s. \sqrt{2\pi}} \cdot \exp\left(-\frac{(\ln(x) - \mu)^2}{2s^2}\right) \qquad (7)$$

with G = 290 000, µ = 3.42 and s = 1.

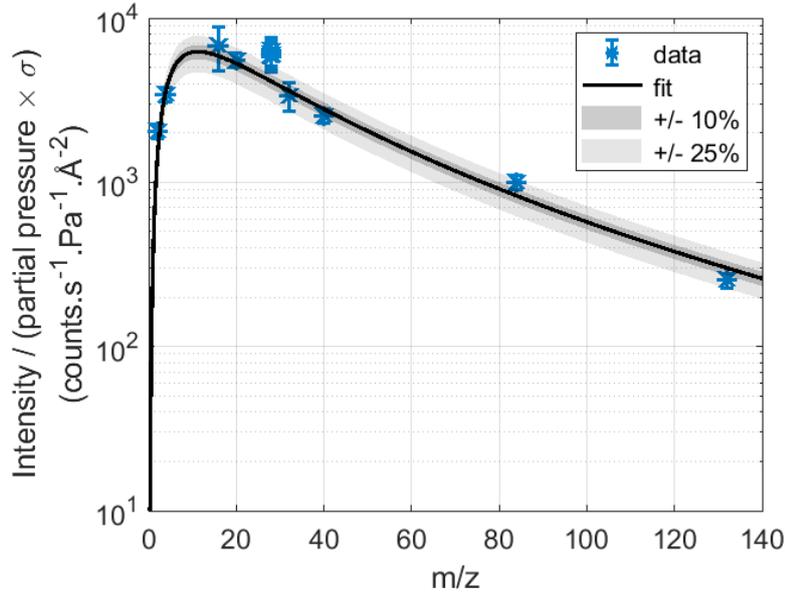

**Figure A4-1.** $T_{tot}(m_X)/\sigma_X$ and its error bars, obtained for a set of parameters tuned on m/z 28. Data points correspond to the following gases at their respective m/z: $H_2$ (2), He (4), $CH_4$ (16), Ne (20), $N_2$ (28), $O_2$ (32), Ar (40), Kr (84) and Xe (132).

Ionization cross sections are hard to measure and model: their values have large uncertainties which affect the calibration. It is especially the case for molecular gases that can fragment. Values given for simple ionization coefficients by different sources are compared in table A4-1: results of computations with the databases of Biagi (program Magboltz) [70] and IST-Lisbon [71,72], and measurements with Phelps database [73–78]. For consistency, we always used Phelps database values when they exist.

| Gas | Simple ionization cross section σ (Å²) | Estimated uncertainty on σ (+/- %) | Isotope corrective coefficient $K_{iso}$ |
|---|---|---|---|
| $H_2$ | 0.935 (Biagi)<br>0.967 (IST)<br>**0.967** (Phelps) | 10 | 1.000 |
| $N_2$ | 2.33 (Biagi)<br>1.92 + 0.23 (IST)<br>**2.15 + 0.17** (Phelps) | 20 | 1.007 |
| $CH_4$ | **1.95** (IST) | 30 | 1.016 |
| $O_2$ | 2.38 (Biagi)<br>2.8 (IST)<br>**2.36** (Phelps) | 20 | 1.005 |
| He | 0.313 (Biagi)<br>0.322 (IST)<br>**0.313** (Phelps) | 10 | 1.000 |
| Ne | 0.490 (Biagi)<br>**0.514** (Phelps) | 10 | 1.105 |
| Ar | 2.77 (Biagi)<br>2.77 (IST)<br>**2.7** (Phelps) | 10 | 1.0035 |
| Kr | **4.21** (Biagi) | 10 | 1.767 |
| Xe | **5.3** (Biagi) | 10 | 3.706 |

**Table A4-1.** Simple ionization cross sections of gases used for calibration.

**Total transmittance for different tunes**

The transmittance of atoms/molecules through the mass spectrometer depends on their mass, but also on the parameters of the MS. The parameters have been tuned once to optimize the intensity of m/z 28 ($N_2^+$), and once to optimize the intensity of m/z 2 ($H_2^+$). Calibration curves have been obtained for both sets of parameters and are plotted in figure A4-2. They give very different shapes, one peaked on m/z ~10, the other on m/z < 2.

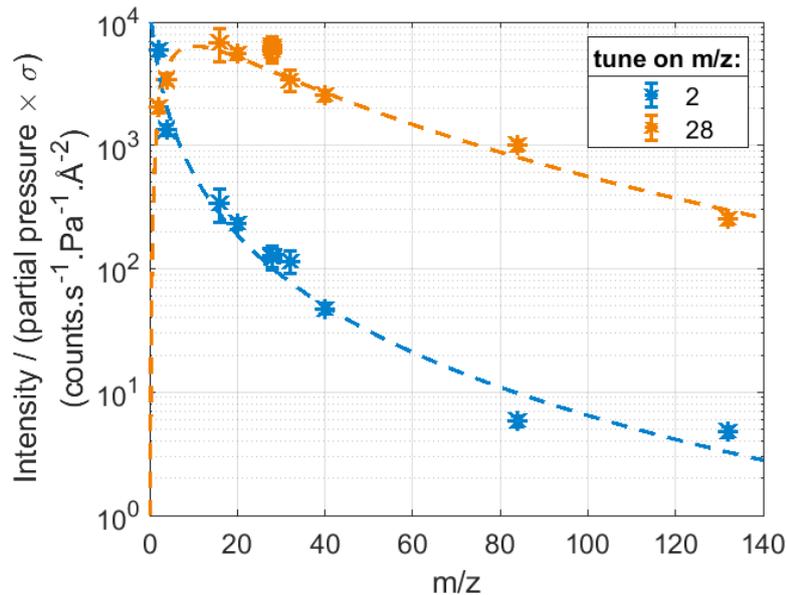

**Figure A4-2**. $T_{tot}(m_X)/\sigma_X$ for two different sets of parameters, obtained for the optimization of the intensity of mass 2 and 28 in a $N_2$-$H_2$ gas mixture. Parameters for the log-normal fit are {G = 37 000, μ = 1 and s = 1.6} for m/z 2 and {G = 290 000, μ = 3.42 and s = 1} for m/z 28.

**Transmittance through the pin hole aperture ($T_{ap}$)**

To obtain the ion flux inside the MS, transmittances through the 100 μm aperture and through the ionization chamber have to be modelled. At the entrance of the MS, the gas goes through a 100 μm hole. The main function of this hole is to reduce the pressure inside the MS, from ~1 mbar to ~$10^{-5}$ mbar. The passage of the hole can have a differential transmittance for the atoms and molecules depending on their mass. It is governed by the flow regime, represented by the Knudsen number, ratio of the Debye length to the characteristic length of the system. The Knudsen number of the system is ~1. Therefore, the flow is in a transition regime between a free molecular flow regime, where gas-wall collisions dominate, and a viscous regime, where gas-gas collisions prevail. Gas flows ($Q$) in the two different regimes are described in [79]:

$$Q_{molecular,X} = A \cdot \frac{v_X}{4} \cdot (P_{X,0} - P_{X,MS}) \quad \text{with } v_X = \sqrt{\frac{8 \cdot k_B T}{\pi \cdot m_X}} \quad [A4.1]$$

$$Q_{viscous,X} = A \cdot P_{X,0} \cdot \sqrt{k_B T} \cdot C_2 \times \frac{1}{\sqrt{m_{avg}}} \quad \text{with } C_2 = C' \times \sqrt{\frac{2\gamma}{\gamma+1}} \times \left(\frac{2}{\gamma+1}\right)^{\frac{\gamma}{\gamma-1}} \quad [A4.2]$$

$$\text{with the respected condition for } [A4.2]: \ \sim 10^{-5} = \frac{P_{MS}}{P_0} \leq \left(\frac{2}{\gamma+1}\right)^{\frac{\gamma}{\gamma-1}} \approx 0.5$$

$A$ is the surface area of the hole. $v_X$ and $m_X$ are the velocity and the mass of the atom / molecule X. $P_{X,0}$ and $P_{X,MS}$ are the partial pressures of X respectively in the reactor chamber and in the MS, with $P_{X,MS} \ll P_{X,0}$. $k_B$ is the Boltzmann constant and T the gas temperature (supposed constant). $m_{avg}$ is the average particle mass of the gas mixture injected. $C'$ is a factor taking into account that high-speed gas stream continues to decrease in diameter after passing through the orifice. For thin circular orifices, $C'$ is equal to ~0.85. $\gamma$ is the specific heat ratio, equals to ~1.4 for diatomic gases, 1.667 for monoatomic gases.

The transition regime is more complex to define. [79] and [80] consider a linear combination of $Q_{molecular,X}$ and $Q_{viscous,X}$, with a coefficient depending on the Knudsen number:

$$Q_{transition,X} = Q_{viscous,X} + Z \times Q_{molecular,X} \quad \text{with } Z \approx \frac{1}{1 + \frac{3\pi}{128} \cdot \frac{1}{Kn}} \quad [A4.3]$$

On the other hand, the gas flow can be expressed as a function of the pumping speed ($S_p$) and the MS pressure [59]:

$$Q_X = P_{MS,X} \times S_p \quad [A4.4]$$

From equations [A4.1], [A4.2], [A4.3] and [A4.4], we can deduce the transmittance through the 100 μm aperture:

$$T_{ap}(m_X) \equiv \frac{P_{X,MS}}{P_{X,0}} \approx \frac{A}{S_p} \cdot \sqrt{k_B T} \times \left(\frac{C_{visc}}{\sqrt{m_{avg}}} + \frac{Z}{\sqrt{2\pi \cdot m_X}}\right) \propto \frac{C_{visc}}{\sqrt{m_{avg}}} + \frac{0.38}{\sqrt{m_X}} \quad [A4.5]$$

with $C_{visc} = 0.46$ for atoms and $C_{visc} = 0.49$ for diatomic molecules and methane.

**Transmittance through the ionization chamber (T$_{ioni}$)**

The ionization source used is based on electronic impact with electrons accelerated at 70 eV. The ionization of species by such an ionization source also leads to the formation of double charged species, and fragments (if there are several atom). Some fragments or double ionized species can be an important percentage of the total ionized species formed. However, it depends strongly on the ionization source used, its tuning and its ageing. Therefore, we avoid to take them into account, and we consider only the simple ionization process on the atom/molecule X: $X + e^- \rightarrow X^+ + e^- + e^-$. Consequently, we use the simple ionization cross section ($\sigma_X$) at 70 eV ($\sigma_X$) to link the partial pressure of the atom/molecule X in the MS ($P_{X,MS}$) to the ion flux formed by ionization ($j_{X^+,MS}$ in m$^{-2}$.s$^{-1}$).

$$\frac{dn_X}{dt} = -\sigma_X n_X n_e v_e \quad \Rightarrow \quad n_X(t) = n_{X,MS} \times \exp(-\sigma_X n_e v_e t) \quad \text{with} \quad t = \frac{L}{v_X} \quad [A4.6]$$

for t corresponding to the crossing of the ionization chamber by X. $L$ is the length of the ionization chamber, $n_X$ (resp. $n_e$) the density of X (resp. electrons) in the ionization chamber, $v_X$ (resp. $v_e$) the velocity of X (resp. electrons). As $\frac{\sigma_X n_e v_e L}{v_X} \sim 10^{-7} \ll 1$, the Taylor expansion gives:

$$n_{X,L} \approx n_{X,MS} \times \left(1 - \frac{\sigma_X n_e v_e L}{v_X}\right) \quad [A4.7]$$

The evolution of X density is directly linked to the formation of the ion X$^+$.

$$n_{X^+,L} = n_{X,MS} - n_{X,L} = n_{X,MS} \times \left(\frac{\sigma_X n_e v_e L}{v_X}\right) \quad [A4.8]$$

Ion flux ($j_{X^+,MS}$) and electron current ($I_e$, in A) are more convenient to use than ion and electron densities in the ionization chamber.

$$j_{X^+,MS} = \frac{1}{4} n_{X^+,L} \times v_{X^+} \quad [A4.9]$$

$$I_e = \frac{1}{4} n_e \times v_e \times e \times l \times L \quad [A4.10]$$

with $v_{X^+}$ the velocity of the ion X$^+$, supposed equal to $v_X$ in the ionization chamber, $e$ the elementary charge, $l$ the width and $L$ the length of the ionization chamber.

The relation between $j_{X^+,MS}$ and $n_{X,MS}$ is obtained by combining [A4.8], [A4.9] and [A4.10]:

$$j_{X^+,MS} = n_{X,MS} \times \sigma_X \times \frac{I_e}{l.e} = P_{X,MS} \times T_{ioni}(X)$$

$$\text{with} \quad T_{ioni}(X) \equiv \frac{j_{X^+,MS}}{P_{X,MS}} = \sigma_X \times \frac{I_e}{k_B.T.l.e} \propto \sigma_X \quad [A4.11]$$

The only dependence in the atom/molecule X chosen is the simple ionization cross section ($\sigma_X$). $T_{ioni}$ has *a priori* no dependence in mass.

# APPENDIX A5. Transmittance of the mass spectrometer for positive ions

**Energy scans of ions**

To compare intensities measured for different ions on a same spectrum, on should check these ions have similar energy distributions, with the same maximum. This comparison has been done on $N_2$-$H_2$ plasma ions. For absolute values of the extractor below ~60 V, and m/z above ~5-10, this condition is valid (see figure A5-1).

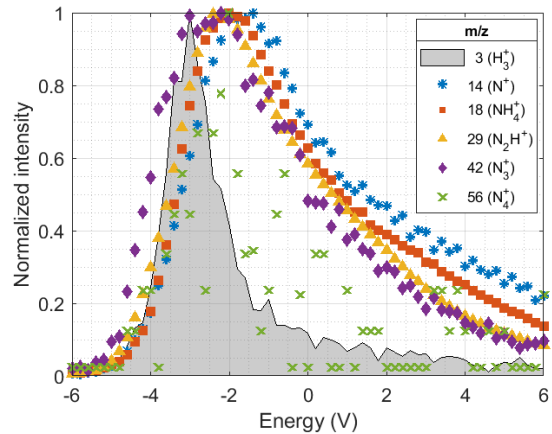

**Figure A5-1**. Energy scans for different ions in a $N_2$-$H_2$ plasma at 0.80 mbar. The extractor was fixed to -40 V. Note: the reference of energy axis is not calibrated and is certainly shifted of a few volts.

**Transmittance of the ions through the group extractor + lens 1**

For absolute values of the extractor below ~60 V, the relation between the extractor and lens 1 is linear, and the scans of lens 1 for different m/z superimpose, except for lower m/z (see figure A5-2). In conclusion, for absolute values of the extractor below 60 V, one should reasonably say that lens 1 focuses similarly all the ions (< 20-30% error) except the ions with low masses (below m/z ~ 5-10).

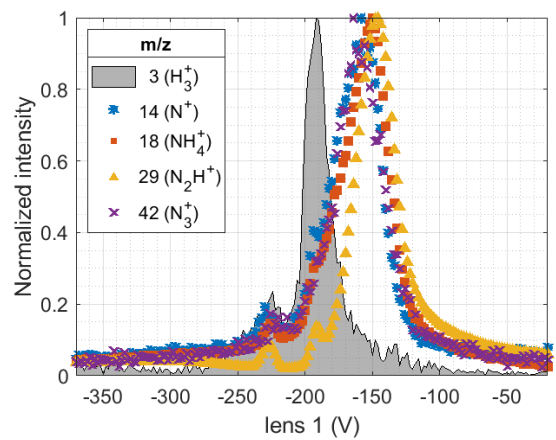

**Figure A5-2.** Scans of the MS parameter lens 1 for different ions, in a $N_2$-$H_2$ plasma at 0.80 mbar, the extractor being fixed to -40 V.